\documentclass[pra,a4paper,twocolumn,superscriptaddress,longbibliography,nofootinbib]{revtex4-2}
\usepackage[utf8]{inputenc}
\usepackage{amssymb}
\usepackage{amsmath}
\usepackage{amsfonts}
\usepackage{amsthm}
\usepackage{amsbsy}
\usepackage{bm}
\usepackage{bbold}

\usepackage{graphicx}
\usepackage{xcolor}
\usepackage{multirow}
\usepackage{float}
\usepackage{comment}
\usepackage{ulem}
\usepackage{relsize}
\usepackage{cmap}
\usepackage{physics}
\usepackage[colorlinks]{hyperref}

\DeclareMathOperator{\sgn}{sgn}

\DeclareMathOperator{\im}{Im}
\DeclareMathOperator{\re}{Re}
\DeclareMathOperator{\arcsinh}{arcsinh}
\DeclareMathOperator{\arccosh}{arccosh}
\DeclareMathOperator{\arctanh}{arctanh}

\begin{document}
	
	\title{Broadened Yu-Shiba-Rusinov states in dirty superconducting films and heterostructures}
	
	\author{S. S. Babkin}
	
	\affiliation{Moscow Institute of Physics and Technology, 141700 Dolgoprudnyi, Moscow Region, Russia}
	\affiliation{\hbox{L.~D.~Landau Institute for Theoretical Physics, acad. Semenova av. 1-a, 142432 Chernogolovka, Russia}}
	
	\author{A. A. Lyublinskaya}
	
	\affiliation{Moscow Institute of Physics and Technology, 141700 Dolgoprudnyi, Moscow Region, Russia}
	\affiliation{\hbox{L.~D.~Landau Institute for Theoretical Physics, acad. Semenova av. 1-a, 142432 Chernogolovka, Russia}}

	\author{I. S. Burmistrov}
	
	\affiliation{\hbox{L.~D.~Landau Institute for Theoretical Physics, acad. Semenova av. 1-a, 142432 Chernogolovka, Russia}}
	
	\affiliation{Laboratory for Condensed Matter Physics, HSE University, 101000 Moscow, Russia
	}

	\date{\today, v.5 - resubmitted} 
	
	\begin{abstract}
	The interplay of a potential and magnetic disorder in superconductors remains an active field of research for decades.
	Within the framework of the Usadel equation, we study the local density of states near a solitary classical magnetic impurity in a dirty superconducting film. We find that a potential disorder results in broadening of the delta-function  peak \color{black} in the local density of states at the Yu-Shiba-Rusinov (YSR) energy. This broadening is proportional to the square root of a normal-state spreading resistance of the film. We demonstrate that
	modification of multiple scattering on the magnetic impurity due to intermediate scattering on surrounding potential disorder \color{black}
	affects crucially a profile of the local density of states in the vicinity of the YSR energy. In addition, we find that 
	a scanning-tunneling-microscopy tip
	can mask an YSR feature in the local density of states. Also, we study the local density of states near a chain of magnetic impurities situated in the normal region of a dirty superconductor/normal-metal junction. We find a resonance in the local density of states near the YSR energy. The energy scale of the resonant peak is controlled by the square root of the film resistance per square in the normal state.
	\end{abstract}

	\maketitle
	
	\section{Introduction}

Studies of the effect of imperfections on superconducting properties have been remaining an active field of research since the middle of the last century. Initially, it was believed that the potential scattering in $s$-wave superconductors does not
affect superconducting properties (so-called, Anderson theorem) \cite{AG0a,AG0b,Anderson1959}. Later it was understood that significant amount of potential disorder results in superconductor to insulator transition \cite{Haviland1989} which is manifestation of competition between Anderson localization and Cooper-channel attraction (see Refs. \cite{gantmakher10,sacepe2020,Burmistrov2021} and references therein).

Classical magnetic impurities being a source for time-reversal symmetry violation cause much severe effect on s-wave superconductivity than potential imperfections.
Without any quantum interference effects taken into account (mean-field approximation), magnetic impurities suppress the superconducting state provided their concentration is high enough \cite{AG2,Skalski1964}. Beyond the Born approximation, the scattering of quasiparticles by a magnetic impurity leads to the appearance of subgap Yu-Shiba-Rusinov (YSR) states in a superconductor \cite{Yu,Soda,Shiba,Rusinov}. 
At a finite concentration of magnetic impurities, YSR states are hybridized and can form  energy bands with hard gaps in the averaged density of states. Depending on the concentration of magnetic impurities and their strength, a rich phase diagram arises (see Ref. \cite{Balatsky} for a review).

Various inhomogeneity effects, such as rare fluctuations of a random potential \cite{LamS-1,LamS-2,MS,MaS}, fluctuations in concentration of magnetic impurities 
 \cite{SilvaIoffe}, fluctuations of superconducting order parameter \cite{LO}, etc. lead to smearing of hard gaps in the density of states (see Refs. \cite{SkF,FS2016} for a review). Recently, it has been shown \cite{BurmistrovSkvortsov2018} that mesoscopic  (point-to-point) \color{black} fluctuations  of effective exchange interaction between spins of magnetic impurity and quasiparticles \color{black} caused by non-magnetic disorder result in strong modification of the YSR bands in the average density of states in comparison with the mean-field analysis. 

For a long time, modification of the superconducting state by a single magnetic impurity has been remaining theoretical concept only \cite{Flatte1997,Flatte1997b}. Progress in scanning tunneling microscopy (STM) makes possible to resolve spatial and energy dependence of YSR states \cite{Yazdani1997,Ji2008,Ji2010,Menard2015,Ruby2016,Choi2017,Perrin2020,Huang2021}. Recent STM experimental studies have revealed rich physics of YSR states in superconductors (see Ref. \cite{Heinrich2018} for a review).    

Currently, experimental studies of solitary YSR states are limited to relatively clean superconductors (typically, Mn or Cr atoms in Pb film or monolayer). Nevertheless, there is an intriguing  and yet unresolved \color{black} question of how non-magnetic disorder affects spatial and energy dependence of YSR states. 
This question can be of additional importance due to the presence of intrinsic magnetic imperfections in nominally non-magnetic disordered superconducting films \cite{Tamir2021}.

Recently, the effect of a random potential on the YSR state has been theoretically studied in Ref.  \cite{Kiendl2017}.  
The authors extended  the scattering approach used in Ref. \cite{Rusinov} 
to incorporate additional scattering on the nonmagnetic impurities. 
\color{black} The broadening of YSR state has been estimated within the lowest order perturbation theory in potential disorder. \color{black} However, behavior of the local density of states (LDoS) near a magnetic impurity has not been addressed. 

\begin{figure*}[t]
\centerline{\includegraphics[width=0.45\textwidth]{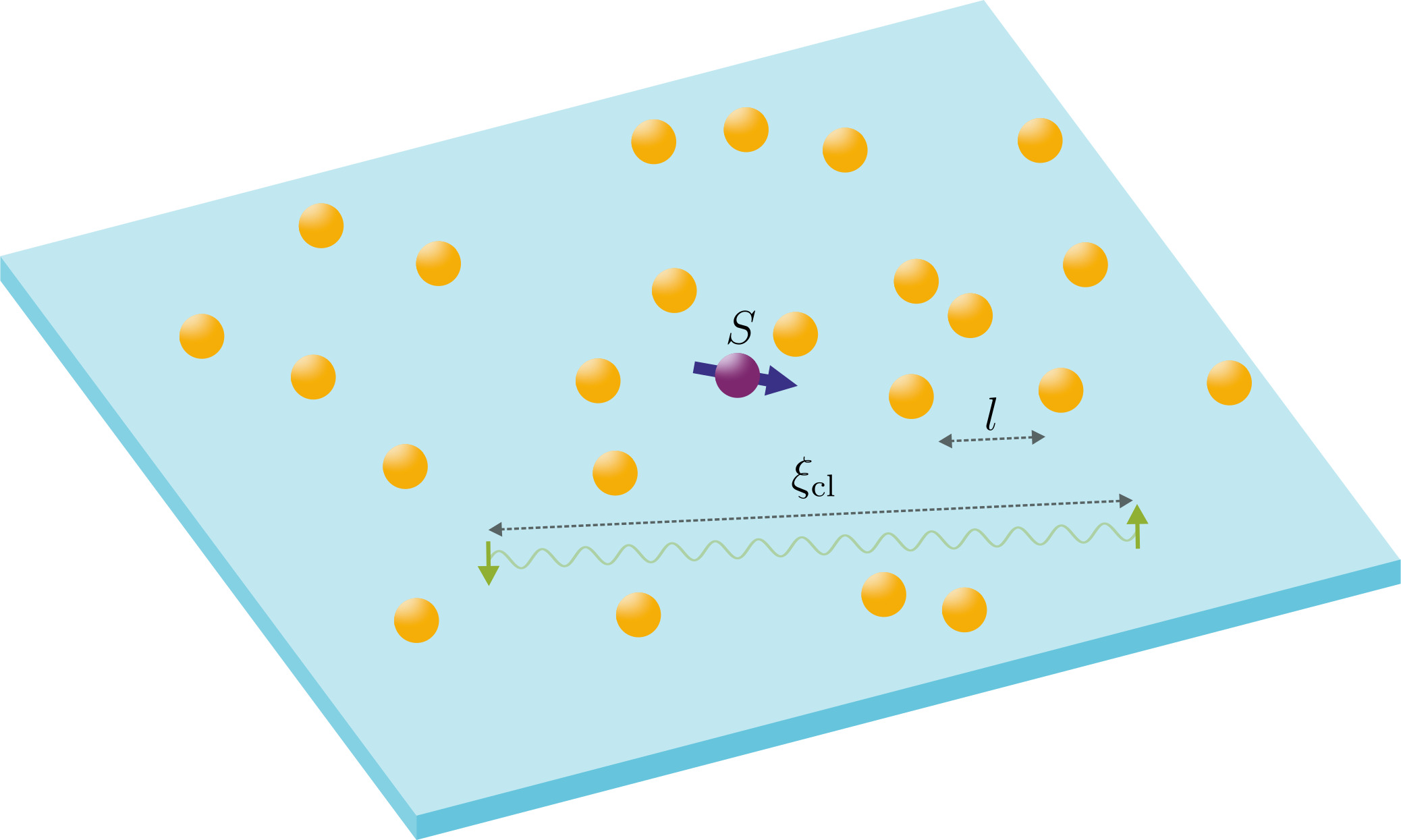}\qquad 
\includegraphics[width=0.45\textwidth]{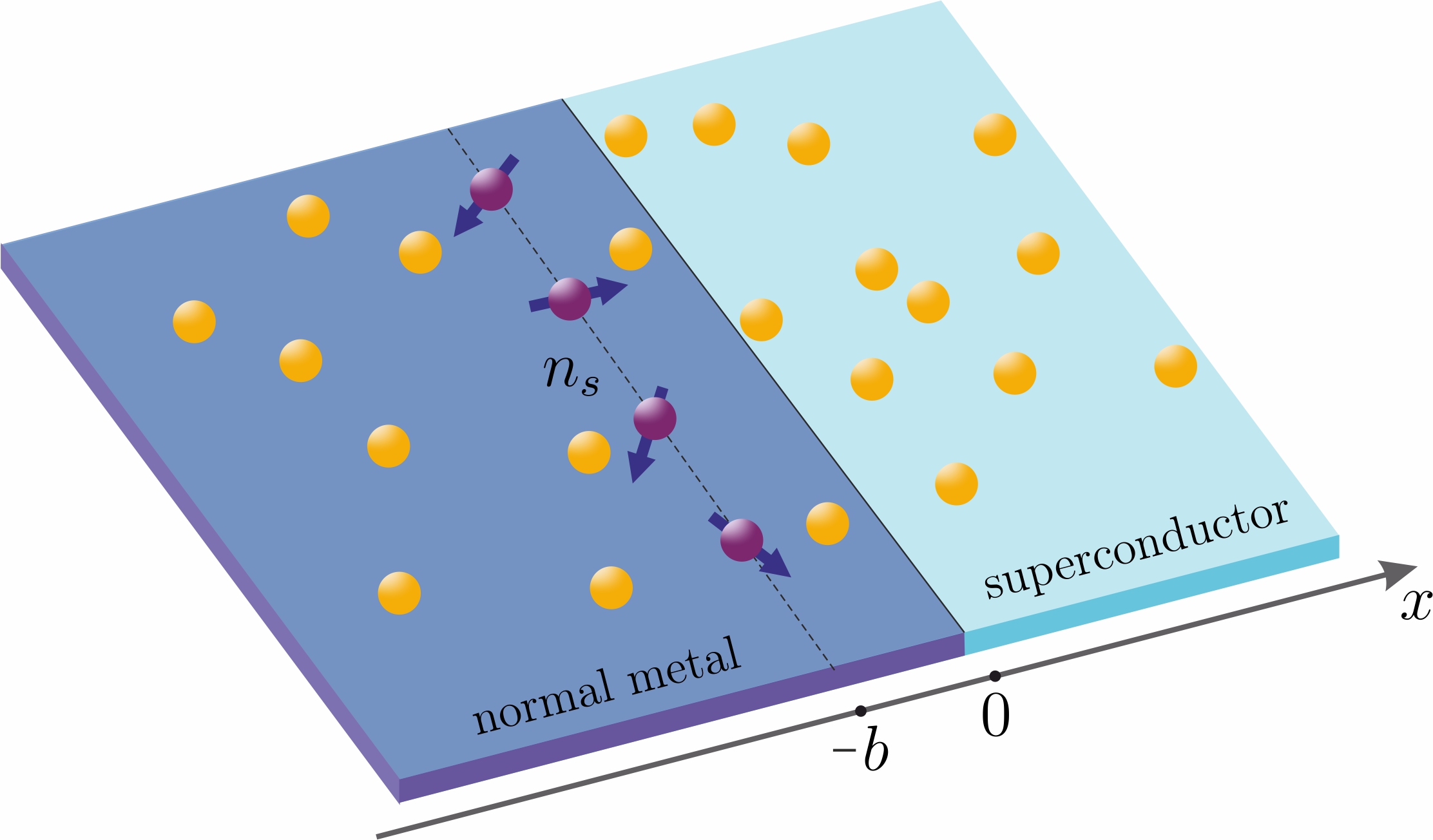}}
\caption{Left: A schematic view of a dirty superconducting film with a solitary magnetic impurity with spin $S$. Potential impurities are shown as yellow circles. A singlet Cooper pair is depicted as a wavy green line. The mean free path is assumed to be much shorter than clean superconducting coherence length, $\ell{\ll}\xi_{\rm cl}$. Right: A schematic view of a dirty superconductor/normal-metal junction, with a chain of magnetic impurities with one-dimensional concentration $n_s$ situated in the normal region at a distance $b$ from the boundary of the materials. Potential impurities are shown as yellow circles.}
\label{Fig:figure1}
\end{figure*} 

Another interesting  question is the fate of YSR states in superconducting heterostructures, e.g., superconductor/normal-metal/superconductor (SNS) or superconductor/normal-metal (SN) junctions. In a clean SNS junction, a magnetic impurity situated in the normal region leads to interesting interplay of YSR states and Andreev levels (see Refs. \cite{Bespalov2018,Bespalov2019} and references therein). We are not aware of similar studies of the LDoS near a magnetic impurity in the normal region of superconducting heterostructures in the dirty regime.

In this paper, we study the LDoS near a solitary classical magnetic impurity in a dirty superconducting film with elastic mean free path, $\ell$, being shorter than the superconducting coherence length, $\xi_{\rm cl}{\gg}\ell$ (see Fig. \ref{Fig:figure1}, left panel). 
Our theoretical analysis is based on the Usadel equation. We investigate the energy and spatial profiles of the LDoS. In the absence of potential disorder, the YSR state due to a single magnetic impurity yields the delta-function contribution to energy 
dependence of the LDoS. We demonstrate that a potential disorder 
results in broadening of
the delta-function into a peak. 
Its energy width is 
controlled by the square root of the spreading resistance of the film  in the normal state\color{black}, see Eq.~\eqref{eq:def:spreading resistance} for the precise definition. 
We find that the profile of the LDoS near the YSR energy is significantly affected by modification of multiple scattering on the magnetic impurity due to intermediate scattering on surrounding potential disorder (treated in the Born approximation). \color{black}
Surprisingly, the corresponding term in the Usadel equation seems to be analogous to the term which takes into account the effect of the mesoscopic fluctuations of the effective magnetic scattering amplitude in the case of finite concentration of magnetic impurities
\cite{BurmistrovSkvortsov2018}. 
\color{black}
 Unexpectedly, \color{black} we find that the potential-disorder-induced broadening of the YSR state at a solitary magnetic impurity  seems \color{black} to be of the order of the variance for the YSR energy
 which is caused by  point-to-point fluctuations 
of the  dimensionless strength of the magnetic impurity
due to the potential disorder
\color{black} found in Refs. \cite{Kiendl2017,BurmistrovSkvortsov2018}. Additionally, we study how the STM tip applied in vicinity of the magnetic impurity masks the YSR feature in the LDoS. 

Also, we investigate the LDoS near a chain of magnetic atoms situated in the normal region of a dirty SN junction (see Fig. \ref{Fig:figure1}, right panel). We find that magnetic impurities increase the LDoS in the vicinity of the YSR energy. However, on the contrary to a homogeneous superconductor, the energy controlling the position of the LDoS peak acquires an imaginary part. The latter means that magnetic impurities in the normal region of the SN junction result in quasibound states rather than the bound ones.     

The outline of the paper is as follows. In Sec. \ref{Sec:1} we calculate the LDoS in a dirty superconducting film with a solitary magnetic impurity. The LDoS in the SN junction with a chain of magnetic atoms is analyzed in Sec. \ref{Sec:2}. The discussion of the obtained results as well as conclusions are given in Sec. \ref{Sec:3}. Some technical details are present in Appendices.

\section{A dirty superconducting film with magnetic impurity\label{Sec:1}}
	
In this section, we consider a dirty superconducting film with a single classical magnetic impurity. We assume that the elastic mean free path $\ell$ is much shorter than the clean superconducting coherence length $\xi_{\rm cl}{=}v_F/\Delta$. Here $v_F$ and $\Delta$ denote the Fermi velocity and the superconducting gap, respectively. We shall treat the problem in the framework of 	
Usadel equation \cite{Usadel1970} which is a standard approach for description of superconductors in dirty limit, $\ell{\ll}\xi_{\rm cl}$. 

\subsection{Standard Usadel equation\label{Sec:1:Stand}}

In the presence of a solitary magnetic impurity situated at the origin of the coordinate system the standard Usadel equation acquires the following form \cite{Fominov2011}, 
\begin{equation}
\frac{D}{2} \nabla^2 \theta_\sigma +i  E \sin \theta_\sigma+\Delta \cos\theta_\sigma
=  \frac{[i\sigma\sqrt\alpha/(\pi \nu)] \sin \theta_\sigma}{1{-}\alpha{+}2i \sigma \sqrt{\alpha} \cos\theta_\sigma} \delta(\bm{r}) \ .
\label{eq:Usadel:1}
\end{equation}
Here  $D$ is the diffusion coefficient in the normal phase, 
 $\sigma{=}\pm$ stands for the projection of an electron spin onto the direction of the impurity spin, and $\delta(\bm{r})$ is the two-dimensional Dirac delta-function. The dimensionless parameter $\alpha{=}(\pi \nu J S)^2$ is the effective strength of the magnetic impurity expressed in terms of the impurity spin, $S$, the exchange interaction constant, $J$, and the density of states at the Fermi level in the normal state (per one spin projection), $\nu$. The spectral angle $\theta_\sigma(E,\bm{r})$ parametrizes the quasiclassical Green's function (see Appendix \ref{App:Usadel}). 
In particular, the spin resolved LDoS is given as
\begin{equation}
\rho_\sigma(E,\bm{r})=\nu \re \cos \theta_\sigma(E,\bm{r}) .
\label{eq:LDOS:1}
\end{equation}

We note that the right-hand side of Eq. \eqref{eq:Usadel:1} is essentially a T-matrix describing  the multiple scattering on the magnetic impurity. \color{black}
Also, we mention that the standard Usadel equation has the following symmetry: the solution $\theta_\sigma$ for the impurity strength $\alpha$ coincides with the solution $\theta_{-\sigma}$ for $1/\alpha$. Therefore, below, when discussing the solution of the standard Usadel equation, we shall consider the case $\alpha{\leqslant}1$. The opposite case, $\alpha{>}1$, can be restored by changing $\sigma$ to $-\sigma$.  

In Eq. \eqref{eq:Usadel:1} we approximate an exchange potential of the magnetic impurity by the delta-function $\delta(\bm{r})$. In fact, the potential has some radius $\lambda$. Since the Usadel equation describes physics at length scales larger than the mean free path, an impurity with $\lambda{\lesssim}\ell$ can be described by the delta-function potential.   
  
It is worthwhile to mention that we neglect the spin-independent part of the potential of the magnetic impurity in Eq. \eqref{eq:Usadel:1}. We shall discuss its effect in Sec. \ref{Sec:3}.

\subsubsection{The LDoS inside the gap, $|E|{<}\Delta$\label{sec:Ldos:e1}}
   
In order to study the LDoS inside the superconducting gap, $E{<}\Delta$, it is convenient to parameterize  the spectral angle as $\theta_\sigma{=}\pi/2{+}i\psi_\sigma$ so that  
\begin{equation}
\rho_\sigma(E,\bm{r})=\nu \im \sinh \psi_\sigma .
\label{eq:LDOS:2}
\end{equation}
In terms of $\psi_\sigma$, the Usadel equation \eqref{eq:Usadel:1} becomes
\begin{gather}
\frac{D}{2} \nabla^2 \psi_\sigma {+} E \cosh \psi_\sigma{-}\Delta \sinh\psi_\sigma
{=} \frac{(\cosh \psi_\sigma)/(2\pi \nu)}{\sigma\sqrt{\beta}+  \sinh\psi_\sigma}\delta(\bm{r})  ,
\notag\\
\beta{=}(1{-}\alpha)^2/(4\alpha) .
\label{eq:Usadel:2}
\end{gather}
In the absence of a magnetic impurity, Eq.~\eqref{eq:Usadel:2} has the homogeneous solution, $\psi_{\infty}{=}\arcsinh (E/\sqrt{\Delta^2{-}E^2})$, corresponding to the density of states in the Bardeen-Cooper-Schrieffer (BCS) theory.  The magnetic impurity disturbs the homogeneous solution, $\psi_\sigma{=}\psi_\infty{+}\delta \psi_\sigma$, where  $\delta \psi_\sigma$ satisfies the two-dimensional sinh-Gordon equation, 
\begin{equation}
\xi^2 \nabla^2 \delta\psi_\sigma - \sinh\delta\psi_\sigma
=  \frac{[\xi^2/(\pi \nu D)] \cosh \psi_\sigma }{\sigma \sqrt{\beta}+ \sinh\psi_\sigma} \delta(\bm{r}) \ .
\label{eq:Usadel:3}
\end{equation}
 Here the length $\xi{\equiv}\xi(E){=}\sqrt{D/(2\sqrt{|\Delta^2{-}E^2|})}$ controls the spatial extent of the perturbation of the homogeneous solution.

As we shall see below, the perturbation $\delta \psi_\sigma$ occurs to be small, $|\delta \psi_\sigma|{\ll} 1$. Then we can approximate the function $\sinh\delta\psi_\sigma$ by its argument in such a way that Eq. \eqref{eq:Usadel:3} reduces to the quantum mechanical problem of a two-dimensional particle in the presence of a delta-function potential (see Appendix \ref{App:Self-consistent}). Therefore, we find    
\begin{equation}
\delta\psi_\sigma(\bm r) = \frac{\tilde{\psi}_\sigma-\psi_\infty}{\ln (\xi/\ell)}K_0(r/\xi), \quad r\geqslant \ell.
\label{eq:sol:deltaPsi}
\end{equation}
Here $K_0(x)$ denotes the modified Bessel function. We note that the mean free path $\ell$ appeared under the logarithm in Eq. \eqref{eq:sol:deltaPsi} as a short-distance regularization for the delta-function. The quantity $\tilde{\psi}_\sigma$ satisfies the following nonlinear algebraic equation,
\begin{equation}
\tilde{\psi}_\sigma=\psi_\infty - \frac{t \sigma \cosh \tilde\psi_\sigma }{\sqrt{\beta} + \sigma  \sinh\tilde\psi_\sigma} .
\label{eq:eq:tilde:psi}
\end{equation}
The perturbation of the homogeneous solution by the impurity is controlled by the parameter
\begin{equation}
t =\frac{2}{\pi g} \ln \frac{\xi}{\ell} ,
\label{eq:def:spreading resistance}
\end{equation}
where $g{=}4\pi \nu D {\equiv} h/(e^2 R_\square)$ is the bare dimensionless normal state conductance of the film. Here $R_\square$ is the resistance per square in the normal phase. We emphasize that the parameter $t$ is energy-dependent, since $\xi$ depends on energy. 

Our approach based on the Usadel equation does not take into account the localization effects. Therefore, our results are limited to the range of energies such that the  $\xi{\ll} \xi_{\rm loc}$ where $\xi_{\rm loc} {\simeq} \ell \exp(\pi g/2)$ is the localization length in two dimensions. This condition is equivalent to  the following inequality,
\begin{equation}
t\ll 1 . 
\label{eq:t:cond}
\end{equation}

\begin{figure}[t]
\centerline{\includegraphics[width=0.45\textwidth]{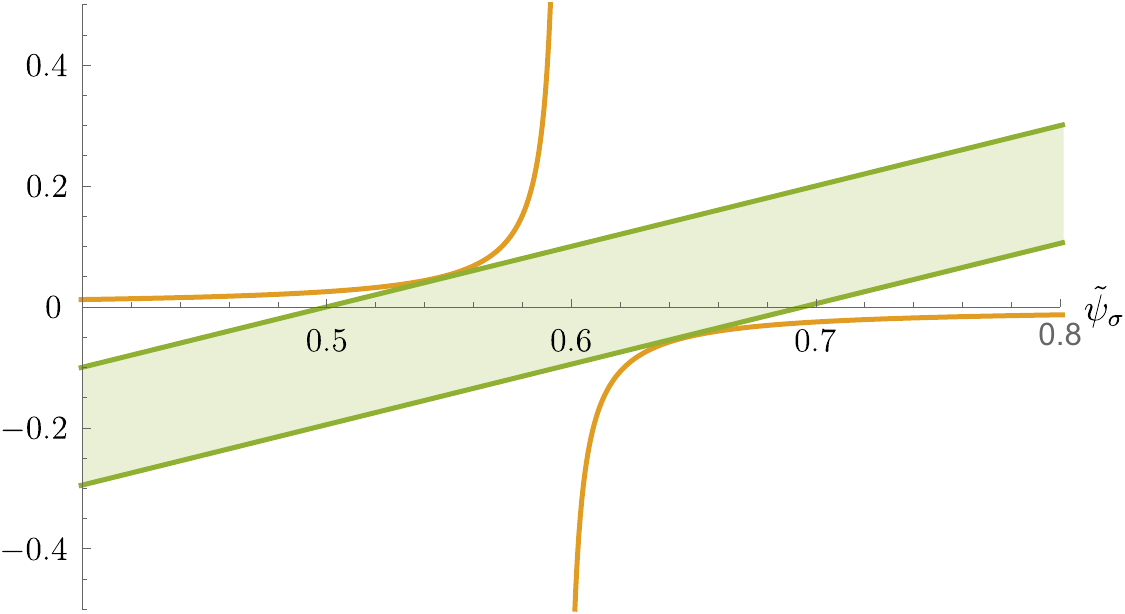}}
\caption{Graphical illustration for Eqs. \eqref{eq:eq:tilde:psi} and \eqref{eq:eq:tilde:psi:bound}. The function ${-}t \sigma \cosh \tilde{\psi}_{\sigma}{/}(\sqrt{\beta}{+}\sigma \sinh \tilde{\psi}_{\sigma})$ is shown in orange color and the function $\tilde{\psi}_{\sigma}{-}\psi_{\infty}$ is shown in green lines that correspond to different values of energy, $E{=}E_{<}$ and $E{=}E_{>}$. The shaded bright-green area corresponds to straight lines $\tilde{\psi}_{\sigma}-\psi_{\infty}$ that do not  cross the orange curve and, therefore, correspond to complex solution for $\tilde{\psi}_{\sigma}$. So, the non-zero density of states appears at energies $E_{<}{<}E{<}E_{>}$.
We use the following values:  $\alpha{=}0.3$, $t{=}0.0025$, and $\sigma{=}{-}1$. For such parameters, the boundary energies are $E_{<}/\Delta{\simeq} 0.46$ and $E_{>}/\Delta{\simeq}0.60$.} 
\label{Fig:figure2}
\end{figure}

In view of the relation \eqref{eq:t:cond} one could try to solve Eq. \eqref{eq:eq:tilde:psi} iteratively, substituting $\psi_\infty$ for $\tilde\psi_\sigma$ in the right-hand side.
However, there are two energies, $E{=}{\pm}E_{\rm YSR}$, where 
\begin{equation}
E_{\rm YSR}=\Delta\sqrt{\frac{\beta}{1+\beta}}=\Delta \frac{1-\alpha}{1+\alpha} ,
\label{eq:YSR:energy}
\end{equation}
at which the denominator $\sqrt\beta{+}\sigma \sinh \psi_\infty$ in the right-hand side of Eq. \eqref{eq:eq:tilde:psi} diverges. We note that $\pm E_{\rm YSR}$ are just the energies of the localized YSR states in a clean superconductor. The divergence of the denominator in Eq. \eqref{eq:eq:tilde:psi} indicates that near the energy $-\sigma E_{\rm YSR}$, the spectral angle $\tilde{\psi}_\sigma$ can be perturbed from the homogeneous solution $\psi_\infty$ parametrically larger than by the term ${\sim}t$. Also, the zero in the denominator implies the existence of complex solution for the spectral angle $\tilde{\psi}_\sigma$ near the energy $-\sigma E_{\rm YSR}$, as illustrated in Fig. \ref{Fig:figure2}. As it follows from Eq. \eqref{eq:LDOS:2}, the complex solution for $\tilde{\psi}_\sigma$ implies the nonzero density of states in some interval of energies around the energy $-\sigma E_{\rm YSR}$. As shown in Fig. \ref{Fig:figure2}, the boundaries of this interval can be found from the combined solution of Eq. \eqref{eq:eq:tilde:psi} and the following equation,
\begin{equation}
\frac{t (1-\sigma \sqrt{\beta} \sinh \tilde\psi_\sigma)}{(\sqrt{\beta}+\sigma\sinh \tilde\psi_\sigma)^2} = 1.
\label{eq:eq:tilde:psi:bound}
\end{equation}
Although, Eq. \eqref{eq:eq:tilde:psi} can be easily solved numerically, it is instructive to discuss its analytical solution using the condition \eqref{eq:t:cond}.

Expanding  $\sinh \tilde{\psi}_\sigma$ ($\cosh \tilde{\psi}_\sigma$) in the denominator (enumerator) of the fraction in the right-hand side of Eq. \eqref{eq:eq:tilde:psi} to the first (zeroth) order in deviation $\tilde{\psi}_\sigma{-}\psi_\infty$, we find 
\begin{equation}
\tilde{\psi}_\sigma \simeq \psi_\infty- \frac{\sigma\sqrt\beta+\sinh\psi_\infty}{2\cosh\psi_\infty} + i \sqrt{t  - \frac{(\sigma\sqrt\beta+\sinh\psi_\infty)^2}{4\cosh^2\psi_\infty}} .
\label{eq:sol:tilde:psi:sigma:1}
\end{equation}
We note that the choice of the sign in front of the square root corresponds to $\im \tilde{\psi}_\sigma{\geqslant} 0$, that guarantees non-negativity of the density of states. 
Using the explicit solution \eqref{eq:sol:tilde:psi:sigma:1}, one can check that the assumption $|\tilde{\psi}_\sigma{-}\psi_\infty|{\ll}1$
is justified in virtue of the inequality \eqref{eq:t:cond}.

Now we can compute the LDoS, see Eq.  \eqref{eq:LDOS:2}. The result reads,
\begin{gather}
\rho_\sigma(E,\bm{r}) =  \frac{\delta\rho_0(-\sigma E)}{\ln (\xi_\beta/\ell)}K_0\bigr(r/\xi_\beta\bigl ) , \quad r\geqslant \ell ,
\notag \\
\delta\rho_0(E) \simeq 
\frac{\nu (1+\beta)^{3/2}}{2\Delta} \re \Bigl [\Gamma^2-(E_{\rm YSR}-E)^2\Bigr ]^{1/2} ,
\label{eq:LDOS:res:1}\\
\Gamma= \frac{2\sqrt{t_\beta}}{1+\beta} \Delta, \quad t_\beta= \frac{2}{\pi g}\ln \frac{\xi_\beta}{\ell} ,
\notag 
\end{gather}
where $\xi_\beta{=}\xi(0)(1+\beta)^{1/4}$.
Therefore, the energy dependence of the density of states has the semicircle shape with the width $2\Gamma$ around the YSR energy $E_{\rm YSR}$. We mention that the square-root energy dependence of  the density of states corresponds to the treatment of the non-magnetic random potential within the self-consistent Born approximation (see Refs. \cite{Hui2015,Kiendl2017} for details). 
 Unexpectedly, \color{black} $\Gamma/2$  coincides with the variance of the YSR energy due to the presence of potential disorder calculated in Ref. \cite{Kiendl2017}. This indicates that the nonzero LDoS around $E_{\rm YSR}$  caused by diffusive motion of quasiparticles around the magnetic impurity 
\color{black} can be thought, physically, 
as a result 
of 
fluctuations of the YSR energy (more precisely, of $\alpha$, see Ref. \cite{BurmistrovSkvortsov2018}) due to  dependence on a realization of \color{black} potential disorder.     

A few remarks are in order here. At first, we note that for $\alpha{\to} 0$ the condition $t_\beta{\ll} 1$ implies that $\alpha{\gg}(\xi(0)/\xi_{\rm loc})^4$, i.e., the result \eqref{eq:LDOS:res:1} is not applicable for extremely weak impurity strengths. 
Secondly, we mention that the perturbation of the LDoS  around the YSR energy contains exactly one fermion state, 
\begin{equation}
\int \limits_{-\Gamma}^\Gamma dE \int d^2\bm{r} \rho_\sigma(E,\bm{r}) = 1/2 .
\end{equation}
At third, there is the critical impurity strength, $\alpha_c$, such that the density of states at the Fermi energy becomes nonzero for $\alpha{>}\alpha_c$. Using Eq. \eqref{eq:LDOS:res:1}, one finds the critical strength as
\begin{equation}
\alpha_c = 1-4\sqrt{t_0} .
\label{eq:alphac}
\end{equation}
At fourth,  we note that the LDoS per spin \eqref{eq:LDOS:res:1} is asymmetric with respect to the chemical potential. 

In Fig. \ref{Fig:figure3} (left panel) we plot the LDoS obtained from the numerical solution of Eq. \eqref{eq:Usadel:3} and compare it with the analytic solution \eqref{eq:LDOS:res:1}. As one can notice, there is hardly any difference between the numerical and analytical solutions. In accordance with the analytical result \eqref{eq:LDOS:res:1}, the  nonzero LDoS region around $E_{\rm YSR}$ broadens with an increase in $\alpha$. We note an interesting nonmonotonous behavior of the total LDoS with energy. For $\alpha_c{<}\alpha{<}(1{+}3\alpha_c)/4$, the total LDoS has three local maxima: at $E{=}{\pm}E_{\rm YSR}$ and at $E{=}0$.
At $\alpha{>}(1{+}3\alpha_c)/4$ only a single maximum at the Fermi energy, $E{=}0$, remains.

\subsubsection{The LDoS outside the gap, $|E|{>}\Delta$}

For description of the effect of the magnetic impurity on the LDoS outside the superconducting gap, $E{>}\Delta$, it is convenient to parameterize  the spectral angle as $\theta_\sigma{=}i\chi_\sigma$. Then the LDoS becomes
\begin{equation}
\rho_\sigma(E,\bm{r}) = \nu \re \cosh\chi_\sigma .
\label{eq:LDOS:2:a}
\end{equation}
In terms of $\chi_\sigma$, the Usadel equation \eqref{eq:Usadel:1} reads,
\begin{equation}
\frac{D}{2i} \nabla^2 \chi_\sigma + E \sinh \chi_\sigma-\Delta \cosh\chi_\sigma
= \frac{[1/(\pi\nu)] \sinh \chi_\sigma}{\sigma \sqrt\beta+ i\cosh\chi_\sigma}
\delta(\bm{r}) \ .
\label{eq:Usadel:2a}
\end{equation}
Without the right-hand side, Eq. \eqref{eq:Usadel:2} has the homogeneous solution, $\chi_{\infty}{=}\sgn E \arcsinh(\Delta/\sqrt{E^2{-}\Delta^2})$ that reproduces the BCS density of states,
\begin{equation}
\rho_0(E) = \frac{\nu |E|}{\sqrt{E^2-\Delta^2}}, \quad |E|>\Delta .
\end{equation}
The magnetic impurity perturbs the homogeneous solution: $\chi_\sigma{=}\chi_\infty {+}\delta \chi_\sigma$, where  $\delta \chi_\sigma$ solves the following equation, 
\begin{gather}
i \xi^2 \nabla^2 \delta\chi_\sigma - \sgn E \sinh\delta\chi_\sigma
=  -
 \frac{(4\sigma\xi^2/g)\sinh \chi_\sigma }{\sqrt\beta +i \sigma \cosh\chi_\sigma} \delta(\bm{r}) \ .
\label{eq:Usadel:3a}
\end{gather}
As we shall see below, the correction $\delta \chi_\sigma$ occurs to be small, $|\delta \chi_\sigma|{\ll} 1$. Then Eq. \eqref{eq:Usadel:3a} can be easily solved,    
\begin{equation}
\delta\chi_\sigma(\bm r) = \frac{\tilde{\chi}_\sigma-\chi_\infty}{\ln (\xi/\ell)}K_0\left (e^{-i\pi \sgn E /4}r/\xi\right ), \quad r\geqslant \ell.
\label{eq:sol:deltaPsi:a}
\end{equation}
The quantity $\tilde{\chi}_\sigma$ satisfies the following nonlinear algebraic equation,
\begin{equation}
\tilde{\chi}_\sigma=\chi_\infty - \frac{i t \sigma\sinh \tilde\chi_\sigma }{\sqrt{\beta} + i \sigma  \cosh\tilde\chi_\sigma} .
\label{eq:eq:tilde:psi:a}
\end{equation}
It is worthwhile to mention the difference between the equations \eqref{eq:eq:tilde:psi} and \eqref{eq:eq:tilde:psi:a}. In the latter case, we seek the solution $\tilde\chi_\sigma$ with nonzero real part. Due to the imaginary unity in the denominator of the fraction in the right-hand side of Eq. \eqref{eq:eq:tilde:psi:a}, the denominator does not vanish for any real $\tilde\chi_\sigma$. 
Considering the inequality \eqref{eq:t:cond}, we can solve Eq. \eqref{eq:eq:tilde:psi:a} iteratively. Substituting $\chi_\infty$ for $\tilde\chi_\sigma$ in its right-hand side, we obtain
\begin{equation}
\tilde{\chi}_\sigma\simeq \chi_\infty - \frac{i t \sigma\sinh \chi_\infty }{\sqrt{\beta} + i \sigma  \cosh\chi_\infty} .
\label{eq:eq:tilde:psi:a:1}
\end{equation}
As one can check, the condition \eqref{eq:t:cond} guarantees the inequality $|\tilde{\chi}_\sigma{-} \chi_\infty|{\ll}1$.
Now, using Eq.  \eqref{eq:LDOS:2:a}, we find the LDoS at $E{>}\Delta$ and $r\geqslant \ell$,
\begin{gather}
\rho_\sigma(E,\bm{r}) =  \rho_0(E)+\re \frac{\delta\rho_0(E)}{\ln (\xi/\ell)}
\left (1 + i\sqrt\beta \frac{\sqrt{E^2-\Delta^2}}{|E|} \right )
\notag \\
\times K_0\left (e^{-i\pi \sigma \sgn E /4}r/\xi\right ) ,
\label{eq:LDOS:res:1:a:00}
\end{gather}
where 
\begin{gather}
\delta\rho_0(E) \simeq 
 -\nu  \frac{t}{1+\beta}
 \frac{\Delta^2 |E|}{(E^2-E_{\rm YSR}^2)\sqrt{E^2-\Delta^2}} .
\label{eq:LDOS:res:1:a}
\end{gather}
We note that $\delta\rho_0(E)$ is the change of the LDoS at the position of the magnetic impurity. The condition $t{\ll}1$ implies that the result \eqref{eq:LDOS:res:1:a:00} is not applicable for  $|E|{-}\Delta{\ll}(\xi(0)/\xi_{\rm loc})^4\Delta$. As we shall see below in Sec. \ref{Sec:II:3}, the region in which the result \eqref{eq:LDOS:res:1:a:00} is not applicable turns to be wider. 
	
Using Eq. \eqref{eq:LDOS:res:1:a:00}, we find
\begin{gather}
\int d^2\bm{r} \delta\rho_\sigma(E,\bm{r}) = 
\frac{\sigma \nu \sqrt\beta}{2\pi(1+\beta)} \frac{\sgn E}{\sqrt{E^2-\Delta^2}}\frac{\Delta^2}{E^2-E_{\rm YSR}^2} .
\label{eq:int:dos:El}
\end{gather}
We note the appearance of $\sgn E$ in Eq. \eqref{eq:int:dos:El}. Therefore, 
integrating the above expression over $|E|{>}\Delta$, we find that there is no change in the number of states for a given spin projection $\sigma$ at $|E|{>}\Delta$.

\subsubsection{Suppression of the order parameter near the magnetic impurity\label{Sec:II:3}}

To study the suppression of the order parameter near the magnetic impurity, it is convenient to rewrite the Usadel equation \eqref{eq:Usadel:1} for the imaginary (Matsubara) $\varepsilon=\pi T(2n+1)$, rather than real, energies,
\begin{equation}
\frac{D}{2} \nabla^2 \theta_\sigma - |\varepsilon| \sin \theta_\sigma+\Delta \cos\theta_\sigma
=  \frac{[i/(2\pi \nu)] \sin \theta_\sigma}{\sigma \sqrt\beta+i \cos\theta_\sigma} \delta(\bm{r}) \ .
\label{eq:Usadel:1:M}
\end{equation}
The superconducting order parameter satisfies the self-consistent equation, 
\begin{equation}
\Delta(\bm{r}) = \pi T |\gamma_{c0}| \sum_{\sigma=\pm} \sum_{\varepsilon>0} \sin \theta_\sigma(\bm{r}) ,
\label{eq:selfcons} 
\end{equation}
where $\gamma_{c0}{<}0$ is the bare dimensionless attraction interaction in the Cooper channel.

In the absence of the magnetic impurity, Eqs. \eqref{eq:Usadel:1:M} and \eqref{eq:selfcons} reduce to the standard self-consistent equation of the BCS theory
for  the homogeneous superconducting order parameter $\Delta_0$,
\begin{equation}
\Delta_0 = 2\pi T |\gamma_{c0}|\sum_{\varepsilon>0}\sin \theta_\infty, \quad \sin\theta_\infty = \frac{\Delta_0}{\sqrt{\varepsilon^2+\Delta_0^2}} .
\label{eq:self-consist:1}
\end{equation}
It is convenient to introduce $\delta \theta_\sigma{=}\theta_\sigma{-}\theta_\infty$ and $\delta \Delta{=}\Delta{-}\Delta_0$. They describe the deviations of $\theta_\sigma$ and $\Delta$ from the homogeneous solutions. As we shall see below, these deviations are small. Therefore, we can linearize Eqs. \eqref{eq:Usadel:1:M} and \eqref{eq:selfcons} as follows:
\begin{gather}
\xi^2_\varepsilon \nabla^2 \delta\theta_\sigma - \delta \theta_\sigma+
\frac{|\varepsilon|\delta \Delta}{\varepsilon^2+\Delta_0^2} 
= \frac{(4i\sigma \xi_\varepsilon^2/g)\sin \theta_\infty}{\sqrt\beta+i \sigma \cos\theta_\infty} \delta(\bm{r}) \ , \notag
\\
\delta \Delta(\bm{r}) = \pi T |\gamma_{c0}|\sum_{\varepsilon>0}\frac{\varepsilon \delta \theta_\sigma(\bm{r})}{\sqrt{\varepsilon^2+\Delta_0^2}} ,
\end{gather}
where $\xi^2_\varepsilon{=}D/(2\sqrt{\varepsilon^2{+}\Delta_0^2})$. We note that the denominator in the right-hand side of the linearized Usadel equation does not turn into zero. Making the Fourier transform from the spatial coordinate $\bm{r}$ to the momentum $\bm{q}$, 
\begin{equation}
\delta \Delta_q = \!\int d^2\bm{r}\, \delta \Delta(\bm{r})\, e^{-i \bm{q}\bm{r}}, \, 
\delta \theta_{\sigma,q} = \!\int d^2\bm{r}\, \delta \theta_\sigma(\bm{r})\, e^{-i \bm{q}\bm{r}}, 
\end{equation}
we find
\begin{equation}
\frac{\delta\Delta_q}{\Delta_0} = -\frac{4\xi(0)^2}{g} \frac{\mathcal{F}_\beta(q\xi(0),T/\Delta_0)}{\mathcal{F}(q\xi(0),T/\Delta_0)} .
\end{equation}
Here the functions $\mathcal{F}_\beta$ and $\mathcal{F}$ are defined as follows:
\begin{gather}
\mathcal{F}_\beta(q\xi(0),T/\Delta_0) =  T \sum_{\varepsilon>0}\frac{1}{1+q^2\xi_\varepsilon^2}\frac{\varepsilon^2\Delta_0/(\varepsilon^2+\Delta_0^2)}{((1+\beta)\varepsilon^2+\beta\Delta_0^2)} , \notag \\
\mathcal{F}(q\xi(0),T/\Delta_0) = T \sum_{\varepsilon>0}
\frac{q^2 \xi_\varepsilon^2+\Delta_0^2/(\varepsilon^2+\Delta_0^2)}{(1+q^2 \xi_\varepsilon^2)\sqrt{\varepsilon^2+\Delta_0^2}} .
\end{gather}
We note that this result coincide with the expression derived previously (see Eq. (B11) in Ref. \cite{FS2016}).

To estimate the effect of the magnetic impurity on the superconductor order parameter, we consider the case of zero temperature. Then,  at $T{=}0$ the summation over $\varepsilon$ in expressions for the functions $\mathcal{F}_\beta(q\xi(0),T/\Delta_0)$ and $\mathcal{F}(q\xi(0),T/\Delta_0)$ can be performed exactly, and we obtain
\begin{equation}
\frac{\delta\Delta_q}{\Delta_0} = -\frac{4\xi(0)^2}{g} \frac{F_\beta(q^2\xi(0)^2)}{F(q^2\xi(0)^2)} ,
\end{equation}
where 
\begin{equation}
F(z) = \frac{1}{4z} - \frac{\sqrt{|1-z^2|}}{2\pi z} 
\begin{cases}
\arccos z, &  z\leqslant 1, \\
(-)\arccosh z, & z>1 ,
\end{cases}
\end{equation}
and
\begin{gather}
F_{\beta}\left(z\right)=\frac{1}{\left(1+\beta\right)z^{2}-1}
\Biggl \{z F(z) - \frac{\sqrt{\beta+1}-\sqrt\beta}{4\sqrt{\beta+1}}
\notag \\
  - \frac{z \sqrt{\beta}}{2\pi} \arctan(1/\sqrt\beta)\Biggr \} .
\end{gather}
The functions $F(z)$ and $F_\beta(z)$ have the following asymptotic behavior at $z{\gg} 1$:
\begin{equation}
F(z)\simeq \frac{\ln(2z)}{2\pi}, \qquad F_\beta(z)\simeq \frac{F(z)}{z(1+\beta)} . \end{equation}
Hence, we find with logarithmic accuracy the modification of the superconducting order parameter at the location of the magnetic impurity,
\begin{equation}
\frac{\delta\Delta(0)}{\Delta_0} \simeq - \frac{t_0}{1+\beta} .
\end{equation}

For $|E|{<}\Delta_0$ the change in the superconducting order parameter produces the purely real correction to the $\delta\psi_\sigma$,
\begin{equation}
\delta\psi_{\sigma,q}^{(\Delta)} = \frac{4\xi(0)^2}{g} \frac{\Delta_0 E/\sqrt{\Delta_0^2-E^2}}{\sqrt{\Delta_0^2-E^2}+q^2\xi(0)^2\Delta_0} \frac{F_\beta(q^2\xi(0)^2)}{F(q^2\xi(0)^2)} .
\end{equation}
Therefore, suppression of the superconducting order parameter does not affect the average LDoS at $|E|{<}\Delta_0$.

For $E{>}\Delta_0$ we compute the Fourier transform of the correction $\delta \chi^{(\Delta)}_{\sigma}(\bm{r})$ as
\begin{equation}
\delta\chi^{(\Delta)}_{\sigma,q} =-\frac{4}{g} \frac{\xi(0)^2}{\sgn E +i q^2\xi^2} \frac{\Delta_0 |E|}{E^2-\Delta_0^2}\frac{F_\beta(q^2\xi(0)^2)}{F(q^2\xi(0)^2)} .
 \end{equation}
Next, we estimate the correction $\delta\chi^{(\Delta)}_{\sigma}$ at the spatial point where the magnetic impurity is situated, 
\begin{equation}
\re \delta\chi^{(\Delta)}_{\sigma}(0)=
 -\frac{1}{\pi g}  \int\limits_0^\infty dz \frac{\Delta_0 E}{E^2-\Delta_0^2 +z^2 \Delta_0^2} \frac{F_\beta(z)}{F(z)}. 
\end{equation}
We note that the integral over $z{=}q^2\xi(0)^2$ is convergent in the ultraviolet. Performing the integration over $z$, we find
\begin{align}
\re \delta\chi^{(\Delta)}_{\sigma}(0) & \simeq 
- \frac{\Delta_0 \sgn E}{\pi g (1+\beta)\sqrt{E^2-\Delta_0^2}}
\notag \\
\times  &
\begin{cases} 
  \pi^2 (1+\beta) F_\beta(0), & |E|-\Delta_0\ll \Delta_0 , \\
 \ln (|E|/\Delta_0), & |E|\gg \Delta_0 .
\end{cases}
\label{eq:Delta:chi:1}
\end{align}
In derivation of the above result we used expansion in $\delta\chi^{(\Delta)}_{\sigma}$. Therefore, Eq. \eqref{eq:Delta:chi:1} is valid for $|\re \delta\chi^{(\Delta)}_{\sigma}(0)| {\ll}1$, i.e., for energies not too close to the unrenormalized gap, $(|E|-\Delta_0)/\Delta_0 {\gg} F_\beta^2(0)/g^2$ .

Using Eqs. \eqref{eq:LDOS:2:a} and \eqref{eq:Delta:chi:1}, we obtain the following correction to the LDoS due to renormalization of the superconducting  order parameter,
\begin{align}
\delta\rho^{(\Delta)}_0(0) & \simeq
- \frac{\nu \Delta_0^2}{\pi g (1+\beta)(E^2-\Delta_0^2)}
\notag \\
\times &
\begin{cases} 
  \pi^2 (1+\beta) F_\beta(0), & |E|-\Delta_0\ll \Delta_0 , \\
 \ln (|E|/\Delta_0), & |E|\gg \Delta_0 .
\end{cases}
\label{eq:deltaLDOS:2a}
\end{align}
Comparing Eqs. \eqref{eq:deltaLDOS:2a} and \eqref{eq:LDOS:res:1:a}, one can check that for $|E|{\gg}\Delta$ the suppression of the superconducting order parameter results in the substitution of $t$ in Eq.  \eqref{eq:LDOS:res:1:a} by $t_0$. Next, near the band edge, $|E|{-}\Delta_0{\ll} \Delta_0$, one can neglect the renormalization of $\Delta$ for $(|E|{-}\Delta_0)/\Delta_0{\gg} \frac{\pi^4 F^2_\beta(0)}{\ln^2(\xi(0)/\ell)}$ only. In the opposite case, $\frac{\pi^4 F^2_\beta(0)}{\ln^2(\xi(0)/\ell)}{\gg}(|E|{-}\Delta_0)/\Delta_0{\gg} (F_\beta(0)/g)^2$, the correction to the LDoS is dominated by the renormalization of the superconducting order parameter. Since in this work we are interested in the behavior of the density of states at energies $|E|{<}\Delta_0$, we shall not study that regime in details.

\begin{figure}[t]
\centerline{\includegraphics[width=0.9\columnwidth]{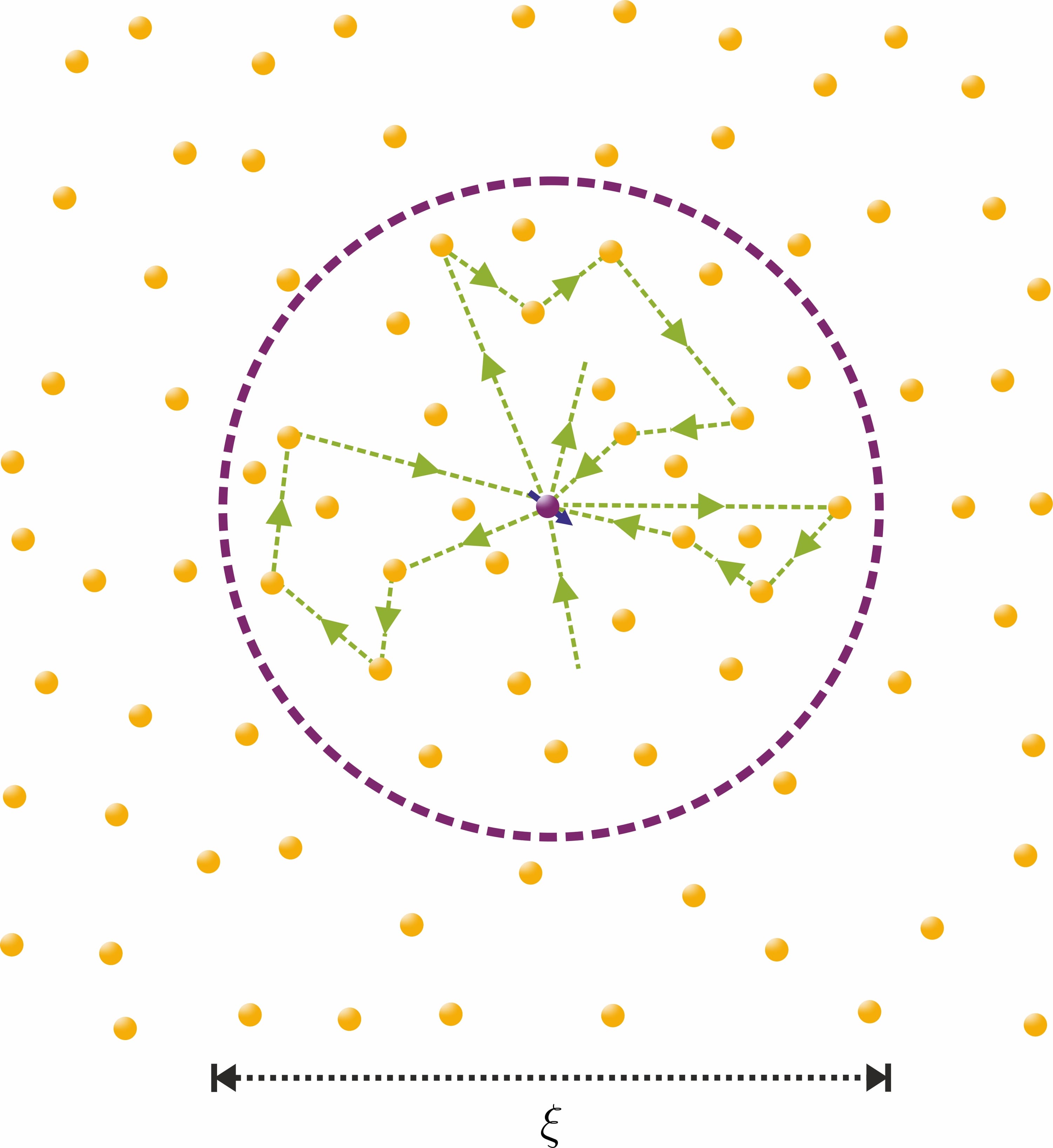}}
\caption{
Sketch of 
quasiparticle scattering on potential disorder between rescattering on the magnetic impurity. Potential impurities are shown as yellow circles, a solitary magnetic impurity with spin $S$ is shown as a purple circle. The schematic trajectory of a quasiparticle is depicted as a dotted green line.} 
\label{Fig:Figure}
\end{figure} 

\begin{figure*}[t]
\centerline{\includegraphics[width=0.45\textwidth]{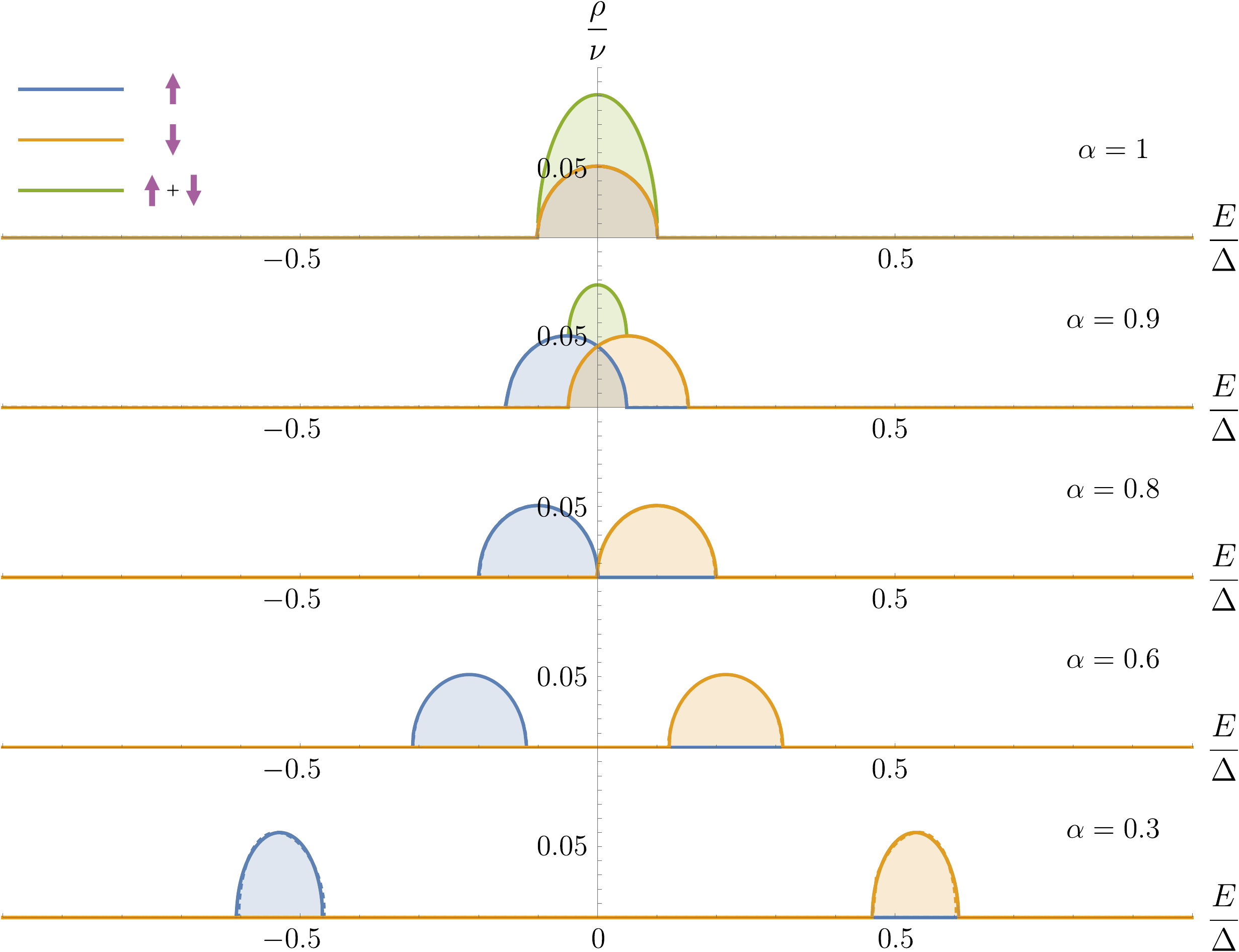}\qquad \includegraphics[width=0.45\textwidth]{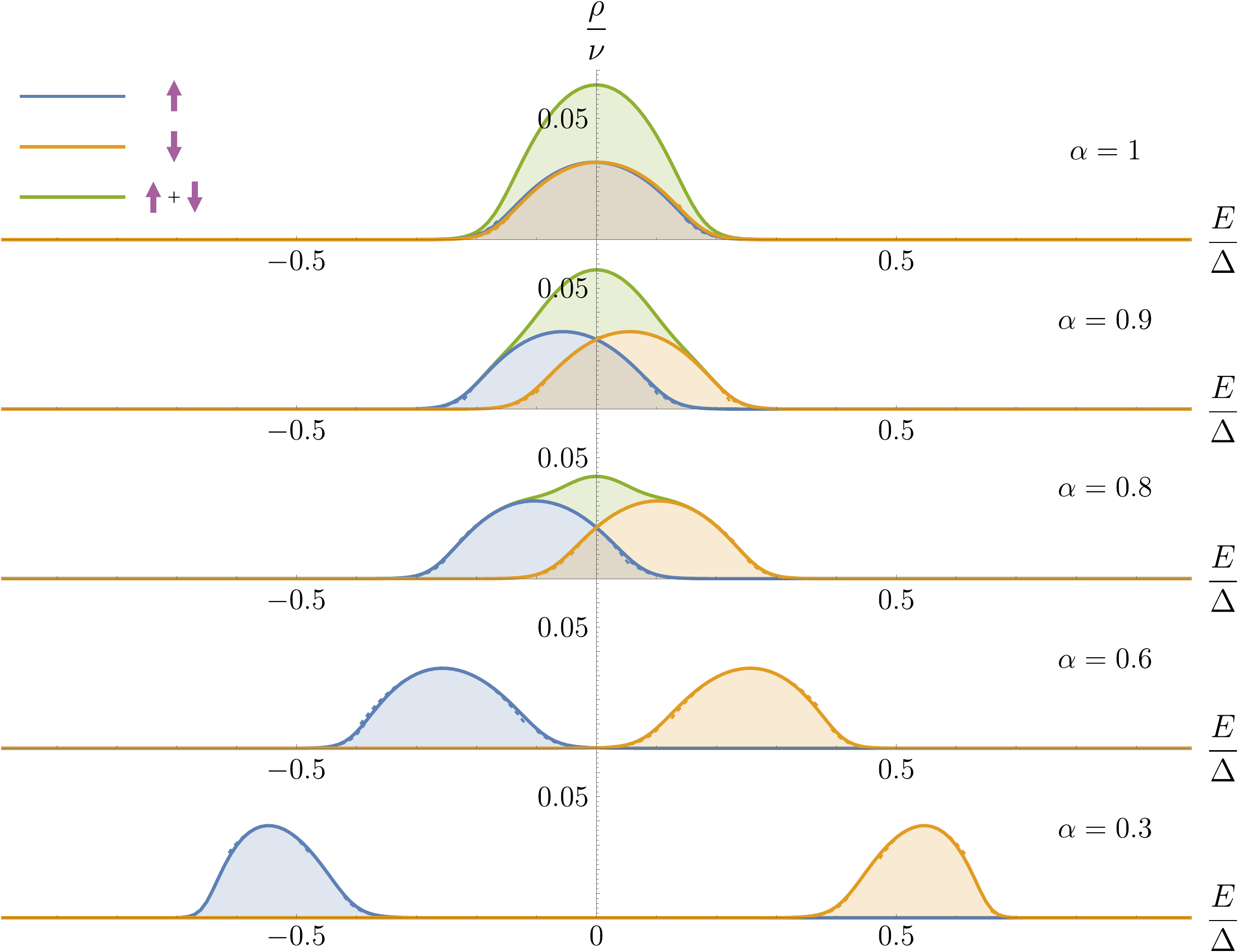}}
\caption{Dependence of the LDoS on energy at the position of the magnetic impurity obtained from standard (left panel) and renormalized (right panel) Usadel equations for different values of $\alpha$. The total LDoS is shown in green, whereas $\rho_\sigma(E,0)$ for $\sigma{=}{-}1({+}1)$ is shown in orange (blue) color. The analytical (numerical) result is depicted as a solid (dotted) line. We choose $t{=}0.0025$ and neglect its weak energy dependence. }
\label{Fig:figure3}
\end{figure*} 

\subsection{Renormalized Usadel equation and the LDoS at $|E|{<}\Delta$}

The solution of the standard Usadel equation \eqref{eq:Usadel:1} results in the broadening of the YSR state due to potential disorder. However, Eq. \eqref{eq:Usadel:1} produces the LDoS with sharp edges, cf. Eq. \eqref{eq:LDOS:res:1}. 
As we discussed above, physically, the broadening of the YSR state can be  understood as the result of 
fluctuations of the impurity strength $\alpha$.  Therefore, one expects smooth energy dependence of the LDoS around the YSR energy. This indicates that the standard Usadel equation \eqref{eq:Usadel:1} is not suited for calculation of the LDoS at $|E|{-}E_{\rm YSR} {\gtrsim} \Gamma$. 

One way to improve the standard Usadel equation is to consider non-symmetric in replica space solutions for the spectral angle, as it was done in Ref. \cite{FS2016}. Here we employ an alternative idea introduced in Ref. \cite{BurmistrovSkvortsov2018} for the case of dilute concentration of magnetic impurities, $n_s^{(2)}$, distributed in the film according to the Poisson distribution. The standard Usadel equation \eqref{eq:Usadel:1} can be derived as the saddle-point of the nonlinear sigma model (NLSM) \cite{MaS}. However, as it was shown in Ref. \cite{BurmistrovSkvortsov2018}, since we are interested in physics at the length scale $\xi$ (alternatively, at the energy scale $E$), we need to renormalize the NLSM action from the mean free path up to $\xi$ (or from elastic scattering rate $1/\tau$ down to $E$). Upon this renormalization, the term describing scattering by magnetic impurities is strongly renormalized. In the case of a single magnetic impurity, there exists similar renormalization of the NLSM such that the renormalized Usadel equation acquires the following form (see Appendix \ref{App:Usadel}),
\begin{align}
\frac{D}{2} \nabla^2 \theta_\sigma + & i  E \sin \theta_\sigma +\Delta \cos\theta_\sigma
\notag \\
& =  \left \langle \frac{[i\sigma\sqrt\textrm{a}/(\pi \nu)] \sin \theta_\sigma}{1{-}\textrm{a}{+}2i \sigma \sqrt{\textrm{a}} \cos\theta_\sigma} \right \rangle_{\textrm{a}}  \delta(\bm{r}) \ .
\label{eq:Usadel:1:mod}
\end{align}
Here $\langle\dots\rangle_{\textrm{a}}$ is the average with respect to the following log-normal distribution,
\begin{equation}
\mathcal{P}_\alpha(\textrm{a},t)=\frac{1}{4\textrm{a} \sqrt{\pi t}} \exp\left [-\frac{1}{4t}\left (\frac{1}{2}\ln\frac{\textrm{a}}{\alpha}+t\right )^2\right ]
\label{eq:P:dist}
\end{equation}
Since $\mathcal{P}_\alpha(\textrm{a},t{\to}0){\to}\delta(\textrm{a}{-}\alpha)$, Eq.  \eqref{eq:Usadel:1:mod} transforms into Eq. \eqref{eq:Usadel:1} as $t{\to} 0$.
 
We note that the right hand side of Eq. \eqref{eq:Usadel:1:mod} can be thought as the T-matrix 
renormalized by scattering of a quasiparticle on potential disorder between rescattering on the magnetic impurity, \color{magenta} this is illustrated schematically in Fig. \ref{Fig:Figure}. 
\color{black}
Surprisingly, the result 
\eqref{eq:Usadel:1:mod} can be obtained from the renormalized Usadel equation of Ref. \cite{BurmistrovSkvortsov2018} upon substitution of $n_s^{(2)}$ by $\delta(\bm{r})$. 

In order to find the LDoS at $|E|{<}\Delta$, we follow the same approach as in Sec. \ref{sec:Ldos:e1}. Parametrizing  the spectral angle as $\theta_\sigma{=}\pi/2{+}i\psi_\sigma$ with $\psi_\sigma{=}\psi_\infty{+}\delta \psi_\sigma$, we find that $\delta \psi_\sigma(\bm{r})$ is given by Eq. \eqref{eq:sol:deltaPsi}, where 
$\tilde{\psi}_\sigma$ satisfies the following nonlinear equation (cf. Eq. \eqref{eq:eq:tilde:psi}), 
\begin{equation}
\tilde{\psi}_\sigma=\psi_\infty - \left \langle \frac{2 t \sigma \sqrt{\textrm{a}} \cosh \tilde\psi_\sigma }{1-\textrm{a}+ 2\sigma \sqrt{\textrm{a}} \sinh\tilde\psi_\sigma} \right \rangle_{\textrm{a}} .
\label{eq:eq:tilde:psi:ren}
\end{equation}
Using the condition \eqref{eq:t:cond}, we rewrite Eq. \eqref{eq:eq:tilde:psi:ren} as
\begin{equation}
\tilde{\psi}_\sigma = \psi_\infty - \frac{\sqrt{t} \sigma \cosh \tilde\psi_\sigma}{\sqrt{1+\beta_t}} \mathcal{H}\left(\frac{\sqrt{\beta_t}+\sigma \sinh \tilde\psi_\sigma}{2\sqrt{t}\sqrt{1+\beta_t}}\right),
\label{eq:eq:H}
\end{equation}
where $\mathcal{H}(z) {=}  \sqrt{\pi} e^{{-}z^2}[\operatorname{erfi}(z){-}i\sgn(\im z)]/2$ and $\beta_t{=} (1{-}\alpha_t)^2/(4\alpha_t)$ with $\alpha_t{=}\alpha e^{{-}2t}$. Although algebraic Eq. \eqref{eq:eq:H} can be solved numerically, it is instructive to discuss its analytic solution in limiting cases. 

We start from the case of the vicinity of the YSR energy, $\bigl||E|{-}E_{\rm YSR}\bigr |{\ll}\Gamma$. In this regime, the argument of the function $\mathcal{H}$ in Eq. \eqref{eq:eq:H} is small. Performing series expansion of $\mathcal{H}$ in its argument and using $|\tilde{\psi}_\sigma {-} \psi_\infty|{\ll} 1$, we find a simple but lengthy result,
\begin{gather}
\tilde{\psi}_\sigma \simeq \psi_\infty\! +2 i z_0 \sqrt{t} \Biggl [ 
1  - \left  (\frac{8(1-h_1) \sqrt{t \beta_t}}{(2+h_1)^2\sqrt{1+\beta_t}}-\frac{i \sigma h_1/z_0}{2+h_1}\right )
\notag \\ 
\times
\frac{\sqrt\beta_t+\sigma\sinh\psi_\infty}{2\sqrt{t}\sqrt{1+\beta_t}}
+ \frac{2ih_2/z_0}{(2+h_1)^3}
\left (\frac{\sqrt\beta_t+\sigma\sinh\psi_\infty}{2\sqrt{t}\sqrt{1+\beta_t}}\right )^2
\Biggr ] .
\end{gather}
Here $h_k{\equiv} \mathcal{H}^{(k)}(iz_0)$ denotes the $k$-th derivative of $\mathcal{H}(z)$ at the point $z_0{\approx} 0.32$. The latter is the positive solution of the equation $z_0 {=} i \mathcal{H}(i z_0)/2$. We note that $h_1{\simeq} 0.59$ and $h_2{\simeq} 0.90 i$. Hence, we obtain the average LDoS in the form of Eq. \eqref{eq:LDOS:res:1} but with 
$\delta\rho_0(E)$ given as ($\bigl  ||E|{-}\tilde{E}_{\rm YSR}\bigr |{\ll}\tilde{\Gamma}$)\footnote{In order to derive this result, we expanded $\sinh\psi_\sigma$ in Eq. \eqref{eq:LDOS:2} to the second order in difference $\delta\psi_\sigma{=}\tilde{\psi}_\sigma{-}\psi_\infty$. We note that still it is enough to use the Usadel equation 
Eq. \eqref{eq:Usadel:3} with a linearized left-hand side.}
\begin{align}
\delta\rho_0(E)
 \simeq & 
 \Biggl [1+t_\beta \frac{1-\alpha}{1+\alpha}
 \left (1 +4 c_1 c_2^2 \frac{1-\alpha}{1+\alpha}\right )\Biggl]
 \notag \\
& \times 
\frac{z_0 \nu (1+\beta)^{3/2} \Gamma}{\Delta}
 \left [ 1 - \frac{(E-\tilde{E}_{\rm YSR})^2}{\tilde{\Gamma}^2} \right ] .  
 \label{eq:DOS:max:ren}
 \end{align}
We note that the typical broadening of the YSR state is enhanced,
\begin{equation}
\tilde\Gamma={\Gamma}/{\sqrt{c_1}},
\end{equation}
where $c_1{=}2 |h_2|/[z_0(2{+}h_1)^3]{\simeq}0.32$. Also, there is a non-zero shift of the energy at which the LDoS has maximum,
\begin{gather}
\tilde{E}_{\rm YSR} \simeq \Delta \left [\frac{1-\alpha}{1+\alpha}
 +\frac{4\alpha t_\beta}{(1+\alpha)^2}\left (1+ 4 c_2\frac{1-\alpha}{1+\alpha} \right )\right ],
 \label{eq:EYSR:mod:UE}
 \end{gather}
where $c_2{=}z_0(4{-}h_1)(2{+}h_1)/(2|h_2|) {\simeq}1.57$.
We emphasize that Eq. \eqref{eq:DOS:max:ren} predicts a dramatic reduction (by a factor of $2 z_0{\approx}0.64$) of the maximal magnitude of the LDoS in comparison with the result \eqref{eq:LDOS:res:1}. 
 We reiterate that \color{black}
the broadening $\tilde\Gamma$ (as well as $\Gamma$)
 is of the order of variance of 
the YSR energy (Eq. \eqref{eq:YSR:energy} with $\alpha$ substituted by
$\textrm{a}$) due to log-normal distribution \eqref{eq:P:dist} of the impurity strength. 
\color{black} 

Now we turn our attention to the study of energy tails in the LDoS. We consider the energy interval in which there are no states within the standard Usadel equation~\eqref{eq:Usadel:3}. In this regime, the argument of the function $\mathcal{H}$ in Eq. \eqref{eq:eq:H} is so large that we can use its asymptotic expression, $\mathcal{H}(z) {\simeq} 1/(2z){-}i [\sgn(\im z)]\sqrt{\pi} \exp({-}z^2)/2$ at $|z|{\gg}1$.
Then, assuming that $\im \tilde{\psi}_\sigma {\ll} 1$, we find,
\begin{gather}
\im \tilde{\psi}_\sigma = \frac{\sqrt{\pi t} \cosh \tilde{\psi}^\prime_\sigma}{2 \sqrt{1+\beta_t}}  \left (1+t \frac{1-\sigma \sqrt{\beta_t}\sinh \tilde{\psi}^\prime_\sigma }{(\sqrt{\beta_t}+\sigma\sinh \tilde{\psi}^\prime_\sigma)^2} \right )
\notag \\
\times \exp \left [-\frac{(\sqrt{\beta_t}{+}\sigma\sinh \tilde{\psi}^\prime_\sigma)^2}{4 t (1+\beta_t)}\right ] .
\end{gather}
Here the quantity $\tilde{\psi}^\prime_\sigma$ is the real part of $\tilde{\psi}_\sigma$, $\tilde{\psi}^\prime_\sigma{\equiv}\re \tilde{\psi}_\sigma$, and satisfies Eq. \eqref{eq:eq:tilde:psi}. Hence, we obtain the LDoS in the parametric form,
\begin{gather}
\delta\rho_0(E) =\nu \frac{\sqrt{\pi t} \cosh^2 x}{2 
\sqrt{1+\beta_t}}  \left (1+t \frac{1+ \sqrt{\beta_t}\sinh x}{(\sqrt{\beta_t}-\sinh x)^2} \right )
\label{eq:DOS:tail:ren2} \\
\times \exp \left [-\frac{(\sqrt{\beta_t}-\sinh x)^2}{4 t (1+\beta_t)}\right ] , \notag \\
\frac{E}{\Delta} = \tanh\Bigl [x - \frac{t \cosh x}{\sqrt{\beta_t}-\sinh x} \Bigr ] .\notag
\end{gather} 
This expression holds for a real variable $x$ that satisfies 
$|\sqrt{\beta_t}{-}\sinh x|{\gg}2\sqrt{t(1{+}\beta_t)}$. This implies that the LDoS is exponentially small away from $E_{\rm YSR}$. In particular, there is finite, albeit exponentially small, ${\sim}\exp[-\beta_t/(4t(1+\beta_t))]$,  LDoS at the Fermi energy, $E{=}0$, for $\alpha{\gtrsim} \alpha_c$.

In Fig. \ref{Fig:figure3} (right panel) we plot the LDoS obtained from the numerical solution of Eq. \eqref{eq:Usadel:1:mod} and compare it against the analytic asymptotes  \eqref{eq:DOS:max:ren} and \eqref{eq:DOS:tail:ren2}. As can be seen, analytics and numerics are in full agreement. The renormalized Usadel equation results not only in suppression of the magnitude of the LDoS near the YSR energy but makes the LDoS to be asymmetric.  
We note that this asymmetry disappears in the total LDoS after merging peaks around $\pm E_{\rm YSR}$. Also, our numerical analysis reveals smaller value of $\alpha_c$ in comparison with Eq. \eqref{eq:alphac}, although we find $1{-}\alpha_c{\sim}\sqrt{t_0}$ still.

\begin{figure}[t]
\centerline{\includegraphics[width=0.99\columnwidth]{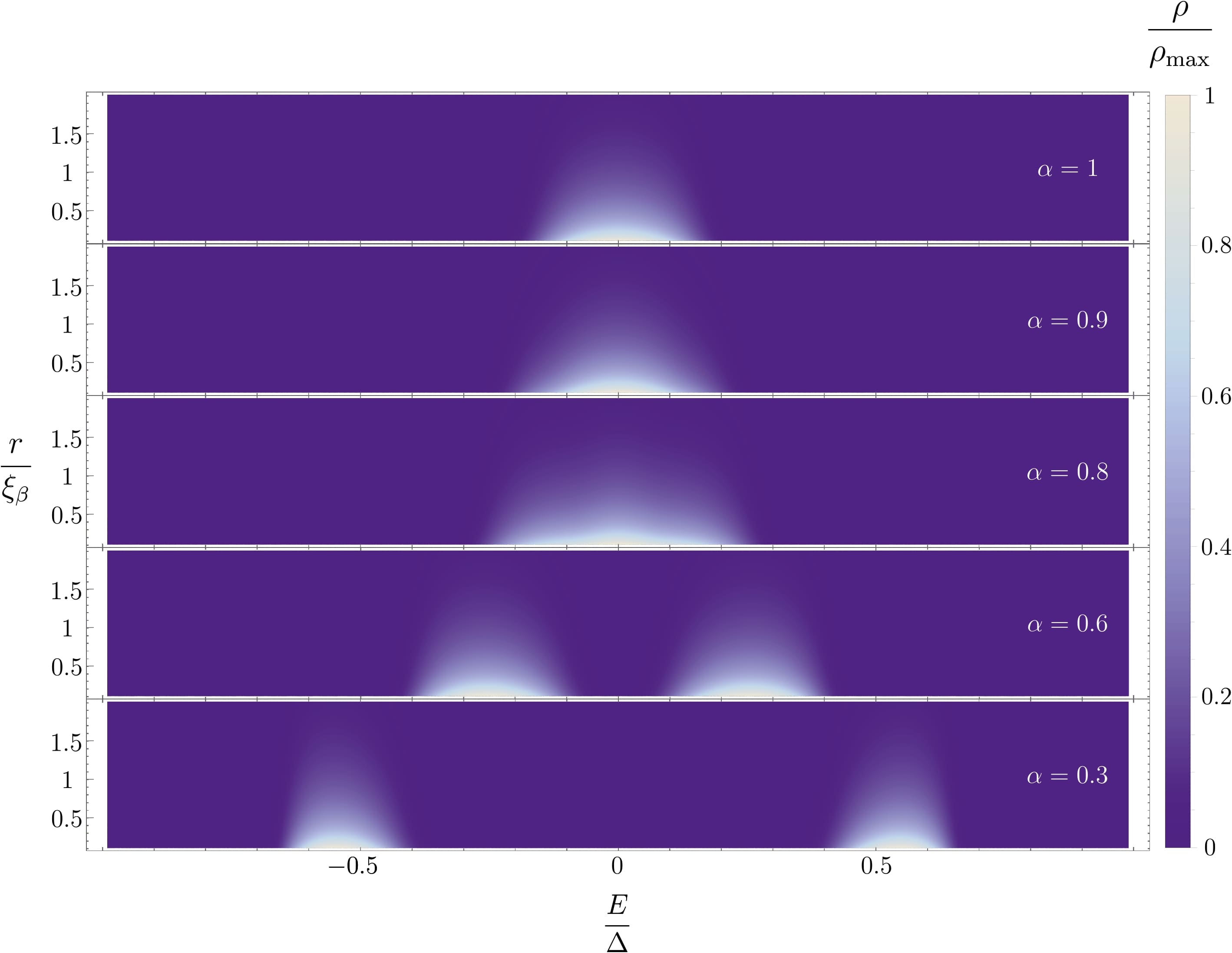}}
\caption{Dependence of the LDoS on energy and distance to the magnetic impurity ($r{\geqslant} l$) obtained numerically from renormalized Usadel equations for different values of $\alpha$. We choose $t{=}0.0025$ and ${l}/{\xi(0)}{=}0.1$.} 
\label{Fig:figure6}
\end{figure}

In Fig. \ref{Fig:figure6} we plot the dependence of the total LDoS on energy and distance to the magnetic impurity for different values of $\alpha$. Here, the LDoS is obtained by the numerical solution of Eq. \eqref{eq:Usadel:1:mod}. For convenience, we normalize the LDoS by its maximal magnitude for each $\alpha$. As can be seen, the LDoS decays with distance on the scale $\xi_\beta$, in full agreement with Eq. \eqref{eq:sol:deltaPsi}.

\subsection{The effect of a tip}

The LDoS in the superconductor can be affected by an STM tip. In this section, we study this effect. We assume that the tip (either superconducting or metallic) is placed near the magnetic impurity. Possibility of tunneling from/to the superconducting film to/from the tip results in modification of the Usadel equation \cite{Skvortsov2000,Skvortsov2001},
\begin{gather}
\frac{D}{2} \nabla^2 \theta_\sigma +i  E \sin \theta_\sigma+\Delta \cos\theta_\sigma
=  \frac{[i\sigma\sqrt\alpha/(\pi \nu)] \sin \theta_\sigma}{1{-}\alpha{+}2i \sigma \sqrt{\alpha} \cos\theta_\sigma} \delta(\bm{r}) 
\notag \\
-\frac{1}{2\pi\nu} \sum_{k=1}^N
\frac{\sqrt{\mathcal{T}_k} \sin(\theta_{\rm tip}-\theta_\sigma)}{1+\mathcal{T}_k+2 \sqrt{\mathcal{T}_k} \cos(\theta_{\rm tip}-\theta_\sigma)}\delta(\bm{r})
\ .
\label{eq:Usadel:1:tip}
\end{gather}
Here we assume $N$ tunneling channels with the tunneling probability $T_k$ each. The quantity  
$\mathcal{T}_k {=}T^2_k/(2{-}T_k)^2$ is the Andreev conductance of the $k$-th channel. For a sake of simplicity, we study the effect of the tip within the standard Usadel equation. As above, we are interested in the modification of the LDoS near the YSR energy alone. 

Following the same steps as in Sec. \ref{Sec:1:Stand}, we find the following equation for the spectral angle at the position of the magnetic impurity, cf. Eq. \eqref{eq:eq:tilde:psi},
\begin{equation}
\begin{aligned}
\tilde{\psi}_\sigma= & \psi_\infty - \frac{t \sigma \cosh \tilde\psi_\sigma }{\sqrt{\beta} + \sigma  \sinh\tilde\psi_\sigma} + C(\tilde\psi_\sigma) , \\
C(\tilde\psi_\sigma) = & \sum_{k=1}^N
\frac{i t \sqrt{\mathcal{T}_k} \cos(\theta_{\rm tip}-i\tilde\psi_\sigma)}{1+\mathcal{T}_k+2 \sqrt{\mathcal{T}_k} \sin(\theta_{\rm tip}-i\tilde\psi_\sigma)} .
\end{aligned}
\label{eq:eq:tilde:psi:tip}
\end{equation}
The term $C(\tilde\psi_\sigma)$ has no singularity in its denominator at the YSR energy. So, the difference between $\tilde{\psi}_\sigma$ and $\psi_\infty$ can be neglected there. Solving Eq. \eqref{eq:eq:tilde:psi:tip} in the same way as Eq. \eqref{eq:eq:tilde:psi}, we find
\begin{align}
\tilde{\psi}_\sigma =& \psi_\infty+\frac{1}{2}C(\psi_\infty)-\frac{\sigma\sqrt{\beta}+\sinh\psi_\infty}{2\cosh \psi_\infty} \notag\\
& +i\sqrt{t-\frac{1}{4}\left[\frac{\sigma\sqrt{\beta}+\sinh\psi_\infty}{\cosh \psi_\infty}+C(\psi_\infty)\right]^2}.
\label{eq:tip:delta:psi}
\end{align}
The above result allows us to compute the LDoS near the energy $E_{\rm YSR}$ for arbitrary magnitude of $\theta_{\rm tip}$. For concreteness, we consider the cases of normal metal and superconducting tips only.

\subsubsection{Normal-metal tip}
 
For a normal-metal tip, the spectral angle is zero, $\theta_{\rm tip}{=}0$. Then, using Eq.
\eqref{eq:tip:delta:psi}, we find that the LDoS is given by the expressions \eqref{eq:LDOS:res:1} but with 
\begin{align}
\delta\rho_0(E) = &  \frac{\nu(1+\beta)^{3/2}}{2\Delta}
\Biggl [ \im \tilde{E}_{\rm YSR}
\notag \\
& + \re \Bigl ( \Gamma^2 - (E-\tilde{E}_{\rm YSR})^2\Bigr )^{1/2}\Biggr ] .
\label{eq:tip:EYSR}
\end{align}

Here we introduced the complex YSR energy,
\begin{equation}
\tilde{E}_{\rm YSR}=E_{\rm YSR} 
\Biggl (1
-\sum_{k=1}^N
\frac{i t\sqrt{\mathcal{T}_k/\beta}}{1+\mathcal{T}_k+2 i \sqrt{\beta\mathcal{T}_k}}
\Biggr ) .
\end{equation}

The normal-metal tip results not only in a shift ${\sim}t$ of the YSR energy, but also the appearance of an imaginary part ${\sim}t$. The latter signals that the YSR state becomes a quasibound one, since it can decay into the normal tip. The existence of the imaginary part in $\tilde{E}_{\rm YSR}$ smears the sharp edges of the LDoS.

\subsubsection{Superconducting tip}

In the case of a superconducting tip with large superconducting order parameter, we can neglect the energy dependence of the spectral angle and use the following approximation: $\theta_{\rm tip}{=}\pi/2$. Then, using Eq. \eqref{eq:tip:delta:psi}, we obtain the LDoS given by Eq. \eqref{eq:tip:EYSR} with
\begin{equation}
\tilde{E}_{\rm YSR}=E_{\rm YSR}
\Biggl (1+
\sum_{k=1}^N\frac{t \sqrt{\mathcal{T}_k/(1+\beta)}}{1+\mathcal{T}_k+2 \sqrt{(1+\beta)\mathcal{T}_k}}
\Biggr) .
\end{equation}

We mention that the superconducting tip results in the shift of the YSR energy. The imaginary part of $\tilde{E}_{\rm YSR}$ is zero due to the absence of quasiparticle tunneling into the superconducting tip. Therefore, sharp edges of the LDoS near the YSR energy remain.  
	
\section{YSR resonance in a dirty SN junction\label{Sec:2}}

Now we discuss how magnetic impurities situated in a dirty normal metal near a superconductor's boundary affect the LDoS. We consider 
a dirty two-dimensional SN junction with a rare chain of magnetic atoms with one-dimensional concentration $n_s$.
For a sake of simplicity, we assume that both the SN boundary situated at $x{=}0$ and the chain of magnetic atoms situated at $x{=}{-}b$ are straight and parallel to each other 
(see Fig. \ref{Fig:figure1}b). Also, we suppose that the spins of magnetic atoms are classical, statistically independent vectors of the length $S$ with the flat distribution over their orientations.

The effect of impurities will be estimated based on the change in the LDoS in comparison with the one without magnetic atoms. We expect that in the presence of a normal metal, the localized state at the YSR energy is smeared out, forming a peak with a finite width. Thus, our goal is to determine the conditions under which the presence of impurities affects the LDoS near the YSR energy most pronounced.

\subsection{Standard Usadel equation\label{III:standard}}

To describe the LDoS in a dirty SN junction with a chain of magnetic impurities, we employ the standard Usadel equation. Contrary to Eq. \eqref{eq:Usadel:1}, the spectral angle is now independent of the spin projection. Due to the heterogeneity of our model, the Usadel equation should be written separately in the regions of the superconductor ($x{>}0$) and the normal metal ($x{<}0$). Under the assumption of an infinite system size in the $y$ direction (see Fig. \ref{Fig:figure1}b), the spectral angle depends solely on the $x$ coordinate. Then the standard Usadel equation becomes
\begin{equation}
\frac{D}{2}\partial_x^2\theta_{\rm s} + iE\sin\theta_{\rm s} + \Delta\cos\theta_{\rm s} = 0
\label{III:eq:Usadel:1a}
\end{equation}
for $x{>}0$ (the superconductor), and
\begin{equation}
\frac{D_{\rm n}}{2}\partial_x^2\theta_{\rm n}+iE\sin\theta_{\rm n}
= \frac{[n_s\alpha/(\pi\nu_{\rm n})] \sin 2\theta_{\rm n}}{1+\alpha^2 +2  \alpha \cos2\theta_{\rm n}} \delta(x+b)
\label{III:eq:Usadel:1}
\end{equation}
for $x{<}0$ (the normal metal).
Here $D_{\rm n}$ denotes the diffusion coefficient of the normal metal, $\nu_{\rm n}$ is the density of states per one spin projection in the normal metal, and 
$\theta_{\rm s}(E,x)$ ($\theta_{\rm n}(E,x)$) stands for the spectral angle in the superconductor (the normal metal). 

The Usadel Eqs. \eqref{III:eq:Usadel:1a}--\eqref{III:eq:Usadel:1} needs to be supplemented with boundary conditions. We assume that away from the SN boundary the superconductor and the normal metal behave as infinite bulk materials, i.e.,
\begin{equation}
\theta_{\rm s}(E,x\to+\infty)=\frac{\pi}{2}+i\psi_{\infty}, \quad \theta_{\rm n}(E,x\to-\infty)=0 .
\label{BC:1}
\end{equation}
At the SN boundary we employ the following boundary conditions \cite{Kuprianov1988,Nazarov1999},
\begin{equation}
\begin{aligned}
\theta_{\rm s}(E,0) &= \theta_{\rm n}(E,0),  \\ 
g \partial_x\theta_{\rm s}(E,x) \Bigl |_{x{=}0} &= g_{\rm n}\partial_x \theta_{\rm n}(E,x) \Bigl |_{x{=}0}
,
\end{aligned}
\label{III:eq:stitching:1}
\end{equation}
where $g_{\rm n}{=}4\pi \nu_{\rm n} D_{\rm n}$ is the conductance of the normal metal.

The LDoS reads, cf. Eq. \eqref{eq:LDOS:1},
\begin{equation}
\rho(E,x)=\begin{cases}
2\nu\re\cos\theta_{\rm s}(E,x),\quad & x\geqslant0 , \\
2\nu_{\rm n}\re\cos\theta_{\rm n}(E,x),\quad & x<0 .
\end{cases}
\label{III:eq:LDOS:1}
\end{equation}

The Usadel equation \eqref{III:eq:Usadel:1} is justified, provided that the magnetic impurities are rare enough, so the scattering of electrons by them can be considered independently. To guarantee such situation, we assume that the following conditions are satisfied, 
\begin{equation}
n_s \ll 
\begin{cases}
g_{\rm n}/\xi_{\rm n}, &  b\gg\xi_{\rm n} ,\\
g/\xi, &  b\ll \min\{\xi_{\rm n},\xi\}, g/\xi\gg g_{\rm n}/\xi_{\rm n}, \\
\max\{g,g_{\rm n}\}/\xi_{\rm n}, & b\ll \min\{\xi_{\rm n},\xi\}, g/\xi\ll g_{\rm n}/\xi_{\rm n} .
\end{cases} 
\label{eq:ns:cond}
\end{equation}
Here the length $\xi_{\rm n}{=}\sqrt{D_{\rm n}/(2|E|)}$ for the normal metal is introduced. We note that this inequality is analogous to the corresponding condition in the case of impurities scattered in the whole two-dimensional plane (see Ref. \cite{BurmistrovSkvortsov2018} and Appendix \ref{App:Condition}).  

Solving the Usadel equations \eqref{III:eq:Usadel:1a}--\eqref{III:eq:Usadel:1} consists in finding solutions in three domains ($x{<}{-}b$, ${-}b{<}x{<}0$, and $x{>}0$) and stitching these solutions at the points $x{=}{-}b$ and $x{=}0$, applying the boundary conditions \eqref{III:eq:stitching:1}. Using the boundary conditions \eqref{BC:1} at spatial infinity, $x{\to}\infty$, we can immediately write the solutions for the spectral angle in the region $x{<}{-}b$,
\begin{equation}
\theta_{\rm n}(E,x)=4\arctan\left [ \exp \left (\frac{x+b}{\xi_{\rm n}}e^{-i\pi \sgn E/4}\right ) \tan\frac{\theta_{\rm b}}{4}\right ],
\label{III:eq:UsadelSolve:1a}
\end{equation}
and in the domain $x{>}0$,
\begin{equation}
\theta_{\rm s}(E,x)=\frac{\pi}{2}+i\psi_{\infty}+4i\arctanh\left(e^{-x/\xi}\tanh\frac{\psi_{\rm 0}}{4}\right) .
\label{III:eq:UsadelSolve:1}
\end{equation}
The constants $\theta_{\rm b}$ and $\psi_{\rm 0}$ have to be determined from the boundary conditions \eqref{III:eq:stitching:1}.

In order to find the solution for the spectral angle on the interval ${-}b{\leqslant}x{\leqslant}0$, we need to write out the first integral of the equation \eqref{III:eq:Usadel:1},
\begin{equation}
\frac{\xi_{\rm n}^2}{2}\left(\partial_x\theta_{\rm n}\right)^2- i\sgn E\cos\theta_{\rm n}
=C ,
\label{III:eq:UsadelInt:1}
\end{equation}
where $C$ is the constant. This allows us to reduce the solution of Eq. \eqref{III:eq:Usadel:1} to the inversion of the incomplete elliptic integral,
\begin{equation}
\frac{x+b}{\xi_{\rm n}}=\int\limits_{\theta_{\rm b}}^{\theta_{\rm n}(E,x)}\frac{d\theta}{\sqrt{2C+2i\sgn E\cos\theta}}.
\label{III:eq:UsadelInt:2}
\end{equation}
Thus, we have the explicit solutions \eqref{III:eq:UsadelSolve:1a}--\eqref{III:eq:UsadelSolve:1} in the regions $x{\geqslant}0$, $x{\leqslant}{-}b$ and the implicit solution \eqref{III:eq:UsadelInt:2} in the interval ${-}b{<}x{<}0$. To obtain the final result for the spectral angle, it remains to determine the values of the constants $\theta_{\rm b}$, $\psi_{\rm 0}$ and $C$ using the boundary conditions \eqref{III:eq:stitching:1}. Hence, we can derive a system of algebraic equations for these coefficients. Boundary conditions at the point $x{=}0$ yield
\begin{equation}
C=-\frac{2g^2\xi_{\rm n}^2}{g_{\rm n}^2\xi^2}\sinh^2\frac{\psi_{\rm0}}{2}-\sgn E\sinh\left(\psi_{\infty}+\psi_{\rm0}\right).
\label{III:eq:UsadelStitching:1}
\end{equation}
Boundary conditions at the point $x{=}{-}b$ lead to 
\begin{gather}
C=2\left(e^{-i\pi\sgn E/4}\sin\frac{\theta_{\rm b}}{2}+\frac{(4 n_s\xi_{\rm n}\alpha/g_{\rm n}) \sin 2\theta_{\rm b}}{1+\alpha^2 +2 \alpha \cos2\theta_{\rm b}}\right)^2\notag \\
-i\sgn E\cos\theta_{\rm b}.
\label{III:eq:UsadelStitching:2}
\end{gather}
Finally, from Eq. \eqref{III:eq:UsadelInt:2} with  $x{=}0$ and the boundary conditions at $x{=}0$, we find the third relation:
\begin{equation}
\frac{b}{\xi_{\rm n}}=\int_{\theta_{\rm b}}^{\frac{\pi}{2}+i\psi_{\infty}+i\psi_{\rm0}}\frac{d\theta}{\sqrt{2C+2i\sgn E\cos\theta}}.
\label{III:eq:UsadelStitching:3}
\end{equation}

Thereby, the solution of the Usadel equation \eqref{III:eq:Usadel:1a}--\eqref{III:eq:Usadel:1} is fully determined by the algebraic system of equations, \eqref{III:eq:UsadelStitching:1}--\eqref{III:eq:UsadelStitching:3}, and by the functions \eqref{III:eq:UsadelSolve:1a}, \eqref{III:eq:UsadelSolve:1}, and  \eqref{III:eq:UsadelInt:2} in the domains $x{<}{-}b$, $x{>}0$, and ${-}b{<}x{<}0$, respectively. Although Eqs. \eqref{III:eq:UsadelStitching:1}--\eqref{III:eq:UsadelStitching:3} can be solved numerically, at first, it is instructive to discuss their analytic solutions in some limiting cases.

\subsection{The LDoS in the case of $b{=}0$}

The algebraic system of equations, \eqref{III:eq:UsadelStitching:1}--\eqref{III:eq:UsadelStitching:3}, can be solved analytically in the case of magnetic impurities situated exactly at the SN boundary, i.e., at $b{=}0$. In the superconductor, $x{>}0$, the spectral angle is given by  Eq. \eqref{III:eq:UsadelSolve:1}. At $x{<}0$ the spectral angle is described by Eq. \eqref{III:eq:UsadelSolve:1a} with $b{=}0$ and 
\begin{equation}
\theta_{\rm b} = \pi/2+i\psi_\infty +i\psi_{\rm 0}.
\label{III:eq:stitching:2a}
\end{equation}
Next, using Eqs. \eqref{III:eq:UsadelStitching:1} and \eqref{III:eq:UsadelStitching:2}, we find the following closed equation for $\psi_{\rm 0}$,
\begin{gather}
i \gamma \sinh\frac{\psi_{\rm0}}{2}+e^{-i\pi\sgn E/4}\sin\left (\frac{\pi}{4}+i\frac{\psi_\infty +\psi_{\rm 0}}{2}\right )
\notag \\
=\frac{(4 i n_s\xi_{\rm n}\alpha/g_{\rm n}) \sinh [2(\psi_\infty +\psi_{\rm 0})]}{1+\alpha^2 -2  \alpha \cosh[2(\psi_\infty +\psi_{\rm 0})]} .
\label{III:eq:stitching:2}
\end{gather}
Here the energy function $\gamma{=}{g\xi_{\rm n}}/(g_{\rm n}\xi)$  is introduced. 
We note that we choose the sign in front of the term proportional to $\gamma$ in Eq. \eqref{III:eq:stitching:2} in such a way that the equation reproduces the known solution for $\psi_{\rm 0}$ in the absence of magnetic impurities, i.e., at $\alpha{=}0$,
\begin{equation}
\psi_{\rm0,0}\equiv\psi_{\rm0}\Bigl|_{\alpha{=}0}=\ln\frac{\gamma+e^{i\pi (1-\sgn E)/4} e^{-\psi_{\infty}/2}}{\gamma+ e^{-i\pi (1+\sgn E)/4} e^{\psi_{\infty}/2}} .
\label{III:eq:AlgebraicSolve:1}
\end{equation}
We mention that the solution of Eq. \eqref{III:eq:stitching:2} for $E{<}0$ can be obtained from the solution for $E{>}0$ through the transformation $\psi_0 {\to} {-} \psi_0^*$. It guarantees that the LDoS is symmetric with respect to $E{=}0$. Therefore, below we shall focus on the case $E{\geqslant}0$.

For a sufficiently low concentration of magnetic impurities, $n_s$, we can assume that the solution to Eq. \eqref{III:eq:stitching:2} is close to the function \eqref{III:eq:AlgebraicSolve:1} due to the smallness of the term proportional to $n_s$. A noticeable effect of that term in Eq. \eqref{III:eq:stitching:2} can be expected if its denominator, $1{+}\alpha^2{-}2\alpha\cosh[2(\psi_{\infty}{+}\psi_{\rm 0})]$, becomes close to zero. As we discussed above, this expression vanishes when $\psi_{\rm 0}{=}0$ and $\psi_{\infty}{=}\arcsinh(\sqrt{\beta})$, corresponding to the YSR energy, $E_{\rm YSR}$, are substituted into it. Combining these ideas, it becomes clear that a peak in the LDoS near the YSR energy is possible if, at first, the concentration $n_s$ is sufficiently low and, secondly, the function \eqref{III:eq:AlgebraicSolve:1} for the energy $E_{\rm YSR}$ is close to zero. Perturbation of the LDoS away from the YSR energy is weak, provided 
\begin{equation}
n_s\xi/g\ll1 .
\label{III:eq:Condition:1}
\end{equation}
We emphasize that this inequality coincides with the inequality given by Eq. \eqref{eq:ns:cond} for $b{\ll}\min\{\xi_{\rm n},\xi\}$. 
The second condition (smallness of  $\psi_{\rm0,0}$ at the YSR energy) means that 
\begin{equation}
 \sqrt{D_{\rm n}/D}\gg (g_{\rm n}/g) \sqrt{(1-\alpha)/(2\alpha)} .
\label{III:eq:Condition:2}
\end{equation}

\begin{figure*}[t]
\centerline{\includegraphics[width=1.0\textwidth]{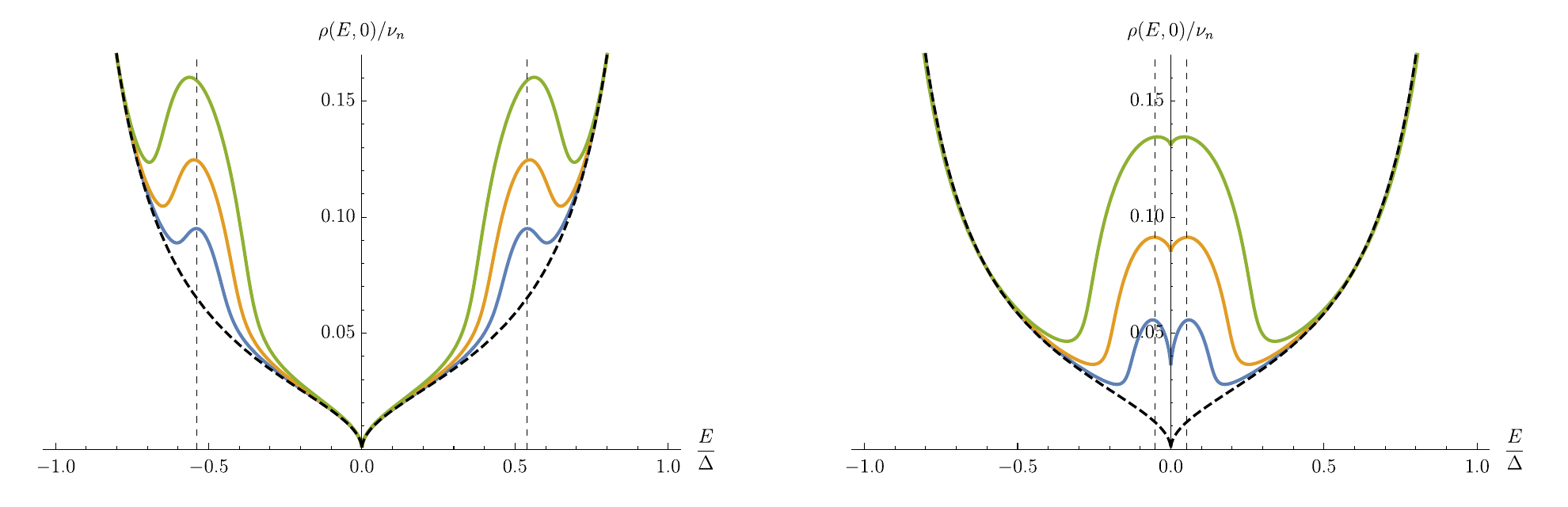}}
\caption{Dependence of the LDoS on the energy at the position of the magnetic impurities in the case of the impurity located at the SN boundary, i.e., at $b{=}0$. Vertical dashed lines denote the position of the YSR energy at a given value of $\alpha$, which is $\alpha{=}0.3$ on the left and $\alpha{=}0.9$ on the right. Blue, orange and green curves show the LDoS for $n_s\xi(0)/g{=}0.001,0.0025,$ and $0.005$, respectively. The dashed curves denote, for comparison, the LDoS without magnetic impurities (i.e., for $n_s{=}0$). We use 
$\sqrt{D_{\rm n}/D}{=}20$ and $g{=}g_{\rm n}$.}
\label{Fig:figure4}
\end{figure*} 

Thus, we assume that, when the conditions \eqref{III:eq:Condition:1} and \eqref{III:eq:Condition:2} are satisfied, the LDoS determined by Eq.~\eqref{III:eq:stitching:2} will coincide with the one given by the solution~\eqref{III:eq:AlgebraicSolve:1} everywhere outside the vicinity of the YSR energy, where the peak is expected. This means that to complete the analytical description of the considered case, $b{=}0$, it is sufficient to solve Eq.~\eqref{III:eq:stitching:2} near the YSR energy. Expanding the left-hand side and denominator of the right-hand side in Eq.~\eqref{III:eq:stitching:2} in small parameter $\psi_{\rm 0}{-}\psi_{\rm 0,0}$  up to the first order, we find
\begin{align}
\psi_{\rm0} =\frac{\psi_{\rm0,0}}{2} & +  \frac{\beta-\sinh^2\psi_\infty}{2\sinh(2\psi_\infty)} 
\notag \\
+ i &\!  \Biggl [ \frac{2 n_s \xi}{g} - \left (\frac{\beta-\sinh^2\psi_\infty}{2\sinh(2\psi_\infty)}-\frac{\psi_{\rm0,0}}{2}\right )^2\Biggr ]^{1/2} .
\label{III:eq:AlgebraicSolve:2}
\end{align}
We note that this result holds for $|\psi_\infty{-}\arcsinh\beta|{\ll} 1$. Next, using Eq. \eqref{III:eq:AlgebraicSolve:2}, we extract the LDoS  
for $\bigl||E|{-}E_{\rm YSR}\bigr|{\ll}{\Delta}/{(1{+}\beta)}$.
In the superconducting region, $x{>}0$, we find
\begin{gather}
\rho(E,x)=\rho_0(E,x)+\delta\rho(E,x).
\label{eq:LDOS:chain:f:r:1}
\end{gather}
Here the first term in the right-hand side is the LDoS in the absence of magnetic impurities,
\begin{equation}
\rho_0(E,x)  \simeq - \frac{2\nu (1+\beta)^{3/2}}{\Delta} e^{-x/\xi_\beta}\im \hat{E}_{\rm YSR} .
\label{III:eq:AlgebraicSolve:S1}
\end{equation}
The second term describes the contribution of magnetic impurities, 
\begin{align}
\delta\rho(E,x) & \simeq \frac{\nu (1+\beta)^{3/2}}{\Delta} e^{-x/\xi_\beta}
\Biggl [\im \hat{E}_{\rm YSR}
\notag \\
& + \re
\sqrt{\Gamma_{\rm n_s}^2-(\hat{E}_{\rm YSR}-|E|)^2}
\Biggr]  .
\label{III:eq:AlgebraicSolve:S2}
\end{align}
Here the energy parameter
\begin{equation}
\hat{E}_{\rm YSR} = \Delta \frac{1-\alpha}{1+\alpha}
\Biggl [1-2^{3/2}\frac{\alpha+i\sqrt\alpha}{(1+\alpha)\sqrt{1-\alpha}}\frac{g_{\rm n}\sqrt{D}}{g \sqrt{D_{\rm n}}} \Biggr ]
\end{equation}
describes the YSR energy modified by the presence of the normal region. We note that, in addition to some shift of the YSR energy in comparison with the case of a homogeneous superconductor, $\hat{E}_{\rm YSR}$ has a negative imaginary part. It indicates that the YSR state becomes the quasibound state rather than the bound one \cite{Bespalov2018}. We mention that for $\alpha{\to}1$ the shift of the YSR energy due to the presence of the SN boundary can become dominant. 

The energy scale 
\begin{equation}
\Gamma_{\rm n_s} = \frac{2 \Delta \sqrt{2n_s \xi_\beta/g}}{1+\beta}  
\end{equation}
determines the effective width of the YSR resonance. We mention that at $n_s{\sim}1/ \xi_\beta$, $\Gamma_{\rm n_s}$ matches the disorder broadening $\Gamma$ for a single magnetic impurity problem, cf. Eq. \eqref{eq:LDOS:res:1}. Also, we note that Eq. \eqref{III:eq:AlgebraicSolve:S2} resembles the result for the LDoS on the magnetic impurity in the presence of the STM tip, cf. Eq. \eqref{eq:tip:EYSR}.

In the normal region, $x{\leqslant}0$, the LDoS can be written in the form of Eq. \eqref{eq:LDOS:chain:f:r:1} with
\begin{align}
\rho_{0}(E,x) \simeq & \nu_{\rm n}\re \cos \Biggl ( 4
\arctan \Biggl ( e^{-|x|\beta^{1/4}(1-i)/(\sqrt{2}\xi_\beta)} \notag \\
\times&
\tan \left (\frac{\pi-i\ln \alpha}{8}\right )\Biggr)\Biggr )
\notag \\
+ & \nu_{\rm n} \frac{g_{\rm n}\sqrt{D}}{g \sqrt{D_{\rm n}}}\frac{1+\alpha}{\sqrt{2\alpha(1-\alpha)}}e^{-|x|\beta^{1/4}/(\sqrt{2}\xi_\beta)}
\notag \\
\times &
\Biggl (\sqrt\alpha \cos\left (\frac{x\beta^{1/4}}{\sqrt{2} \xi_\beta}\right )
+ \sin\left (\frac{x\beta^{1/4}}{\sqrt{2} \xi_\beta}\right )
\Biggr )
\Biggr ]
\label{III:eq:AlgebraicSolve:N1}
\end{align}
and
\begin{align}
\delta\rho(E,x) \simeq & \frac{\nu_{\rm n} (1+\beta)^{3/2}}{\Delta} e^{-|x|\beta^{1/4}/(\sqrt{2} \xi_\beta)} \Biggl [
\cos\left (\frac{x\beta^{1/4}}{\sqrt{2} \xi_\beta}\right )
\notag \\
\times& \Bigl (\im \hat{E}_{\rm YSR}+\re
\sqrt{\Gamma_{\rm n_s}^2-(\hat{E}_{\rm YSR}-|E|)^2}
\Bigr),
\notag \\ 
- &  \sin\left (\frac{x\beta^{1/4}}{\sqrt{2} \xi_\beta}\right )
\Bigl ( 
\re \hat{E}_{\rm YSR}-|E|
\notag\\
-& \im
\sqrt{\Gamma_{\rm n_s}^2-(\hat{E}_{\rm YSR}-|E|)^2}\Bigr )
\Biggr ] .
\label{III:eq:AlgebraicSolve:N2}
\end{align}
We note that according to Eqs. \eqref{III:eq:AlgebraicSolve:N1} and \eqref{III:eq:AlgebraicSolve:N2}, the LDoS in the normal metal oscillates with the distance from the SN boundary (and the magnetic impurities). However, the period of these oscillations coincide with the decay length such that they are not visible.

\begin{figure}[b]
\centerline{\includegraphics[width=0.5\textwidth]{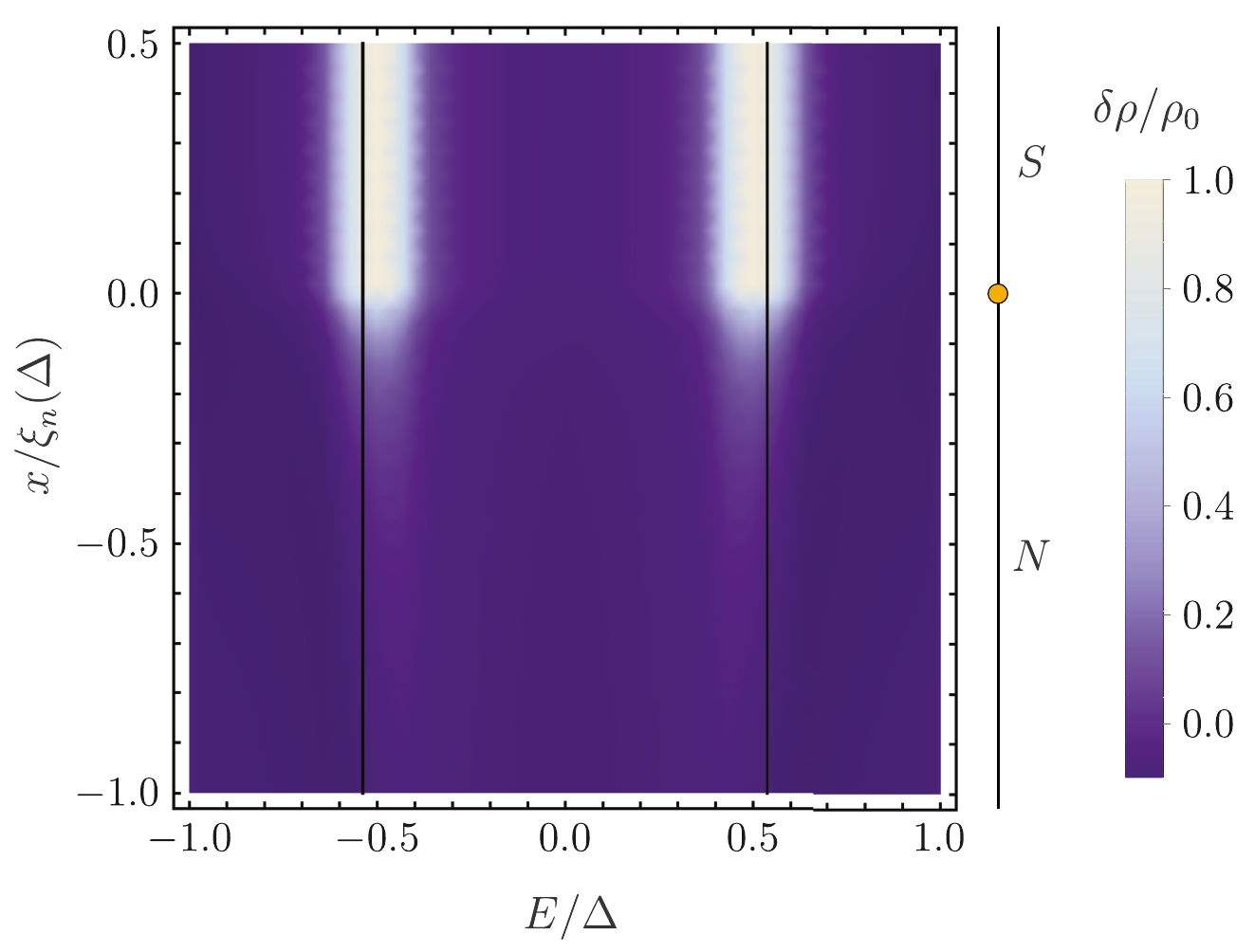}}
\caption{Dependence of the relative change in the LDoS,  $\delta \rho{/}\rho_0$, on energy and $x$-coordinate as a result of the magnetic impurities located at the SN boundary (that is, $b{=}0$). Black vertical lines denote the position of the YSR energy ($\alpha{=}0.3$). We use $n_s\xi(0)/g=0.0025$, 
$\sqrt{D_{\rm n}/D}{=}20$ and $g{=}g_{\rm n}$.}
\label{Fig:figure5}
\end{figure} 

\begin{figure*}[t]
\includegraphics[width=0.47\textwidth]{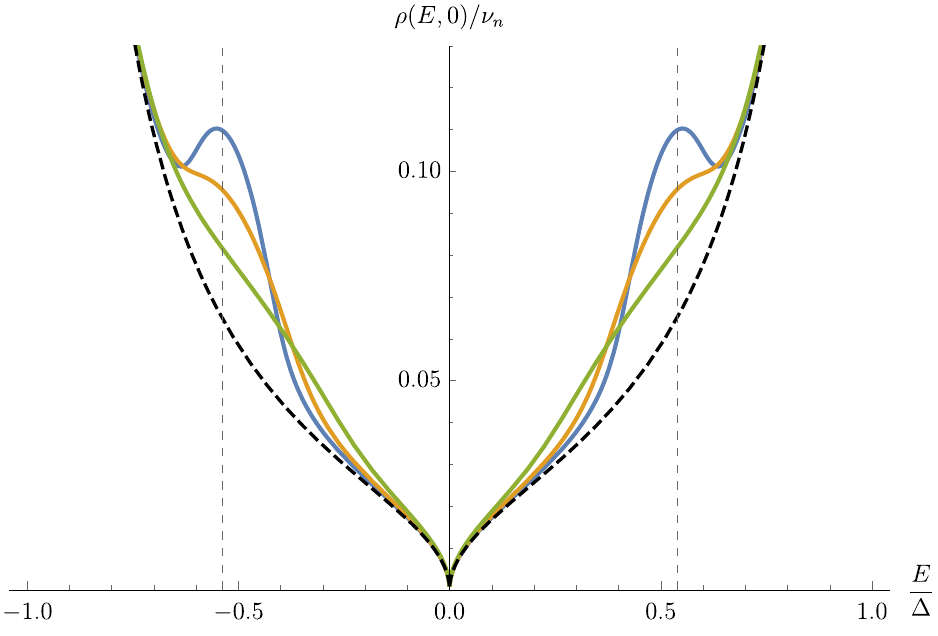}\qquad
\includegraphics[width=0.47\textwidth]{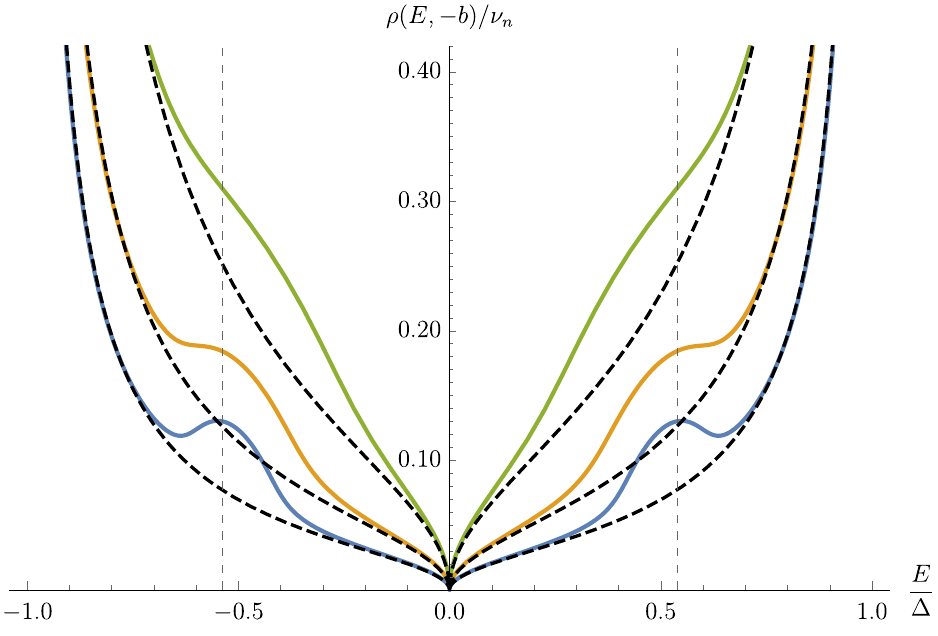}
\caption{Dependence of the LDoS on the energy at the SN boundary (left) and at the position of the magnetic impurities (right). Vertical dashed lines denote the position of the YSR energy at a given value of $\alpha{=}0.3$. Blue, orange and green curves show the LDoS for $b/\xi_{\rm n}(\Delta){=}0.01,0.05,$ and $0.15$, respectively. The dashed curves denote, for comparison, the LDoS without magnetic impurities, i.e., for $n_s{=}0$. We use $n_s\xi(0)/g=0.002$, 
$\sqrt{D_{\rm n}/D}{=}20$ and $g{=}g_{\rm n}$.}
\label{Fig:figure7}
\end{figure*} 

The obtained analytical results are confirmed by a numerical solution. 
 The black dashed curves on graphs in Fig. \ref{Fig:figure4} shows the density of states in the absence of magnetic impurities. We note that V-shape form of the density of states appearing due to inverse proximity effect is reminiscence of the density of states with Thouless energy minigap in the SNS junction. \color{black} The solid curves in Fig. \ref{Fig:figure4} show the sought-for peak in the energy dependence of the LDoS near $E_{\rm YSR}$ for different values of $\alpha$ and $n_s$. The position of the peak is shifted relative to the YSR energy, as predicted in Eq. \eqref{III:eq:AlgebraicSolve:N2}. This picture displays how the peak grows in height and width as the concentration of magnetic impurities $n_{s}$ increases. 
We note that growth of the peak height and width with increasing $n_s$ is limited by the inequality \eqref{III:eq:Condition:1}. Additionally, in the right panel of Fig. \ref{Fig:figure4}, we present how the peaks for positive and negative energies merge when $\alpha$ approaches unity.

Fig. \ref{Fig:figure5} shows the decrease in the relative size of the peak (as before, shifted from $E_{\rm YSR}$) with distance from the impurity and the superconductor. 
As can be seen, the effect of magnetic impurities extends over distances of the order of the length $\xi_{\rm n}$ inside the normal metal. Interestingly, there is no decay with the distance of the relative correction to the LDoS, $\delta\rho{/}\rho_0$, in the superconductor.
This phenomenon can be explained  as follows. In the superconducting part of NS junction ($x>0$) any spatially dependent perturbation decays on the scale of superconducting coherence length $\xi(E)$ for a given energy $E$. There is no other spatial scale within the Usadel equation. In particular, the perturbations of the density of states due to proximity effect in the absence of magnetic impurities at the YSR energy, Eq. \eqref{III:eq:AlgebraicSolve:S1}, and due to magnetic impurities, Eq. \eqref{III:eq:AlgebraicSolve:S2},  decay with the same spatial scale $\xi_\beta$. That is why \color{black} the ratio $\delta\rho/\rho_0$ does not depend on the coordinate, although the magnitude of $\delta\rho$ tends to zero with increasing $x$.

\subsection{The LDoS in the general case, $b{>}0$}

Here we return to the system of Eqs. \eqref{III:eq:UsadelStitching:1}--\eqref{III:eq:UsadelStitching:3}, the solution of which, together with the expressions \eqref{III:eq:LDOS:1}--\eqref{III:eq:UsadelSolve:1}, describes the LDoS for $b{>}0$. As in the previous section, we investigate the region of parameters in which the LDoS has a peak near the YSR energy, and is otherwise close to the impurity-free solution determined by the expressions \eqref{III:eq:stitching:2a} and \eqref{III:eq:AlgebraicSolve:1}.

It is easy to see that the presence of impurities in the system \eqref{III:eq:UsadelStitching:1}--\eqref{III:eq:UsadelStitching:3} is reflected in only one term from the second equation. This term completely coincides with the one investigated in the equation \eqref{III:eq:stitching:2}. Repeating the previous reasoning, we again come to the necessity of fulfilling the inequalities \eqref{III:eq:Condition:1}--\eqref{III:eq:Condition:2}. However, these conditions are not enough. If the impurities are moved to the depth of the normal metal, the proximity effect ceases to work, and the peculiarity in the YSR energy region disappears. For the impurities to remain in the superconductor region of influence, the condition
\begin{equation}
b\ll\xi_{\rm n}
\label{III:eq:Condition:3}
\end{equation}
is necessary. It means that the spectral angle $\theta_{\rm n}$ at the point $x{=}{-}b$ where the magnetic impurities are located should not be close to zero -- its magnitude in a homogeneous normal metal (see \eqref{III:eq:UsadelSolve:1a}).
We mention that condition \eqref{III:eq:Condition:3} becomes more relax with increase of $\alpha$ towards unity. Indeed, at the YSR energy, we find $\xi_{\rm n}(E_{\rm YSR}){=}\xi(0) \sqrt{(1+\alpha)/(1-\alpha)}$.

\begin{figure}[b]
\centerline{\includegraphics[width=0.5\textwidth]{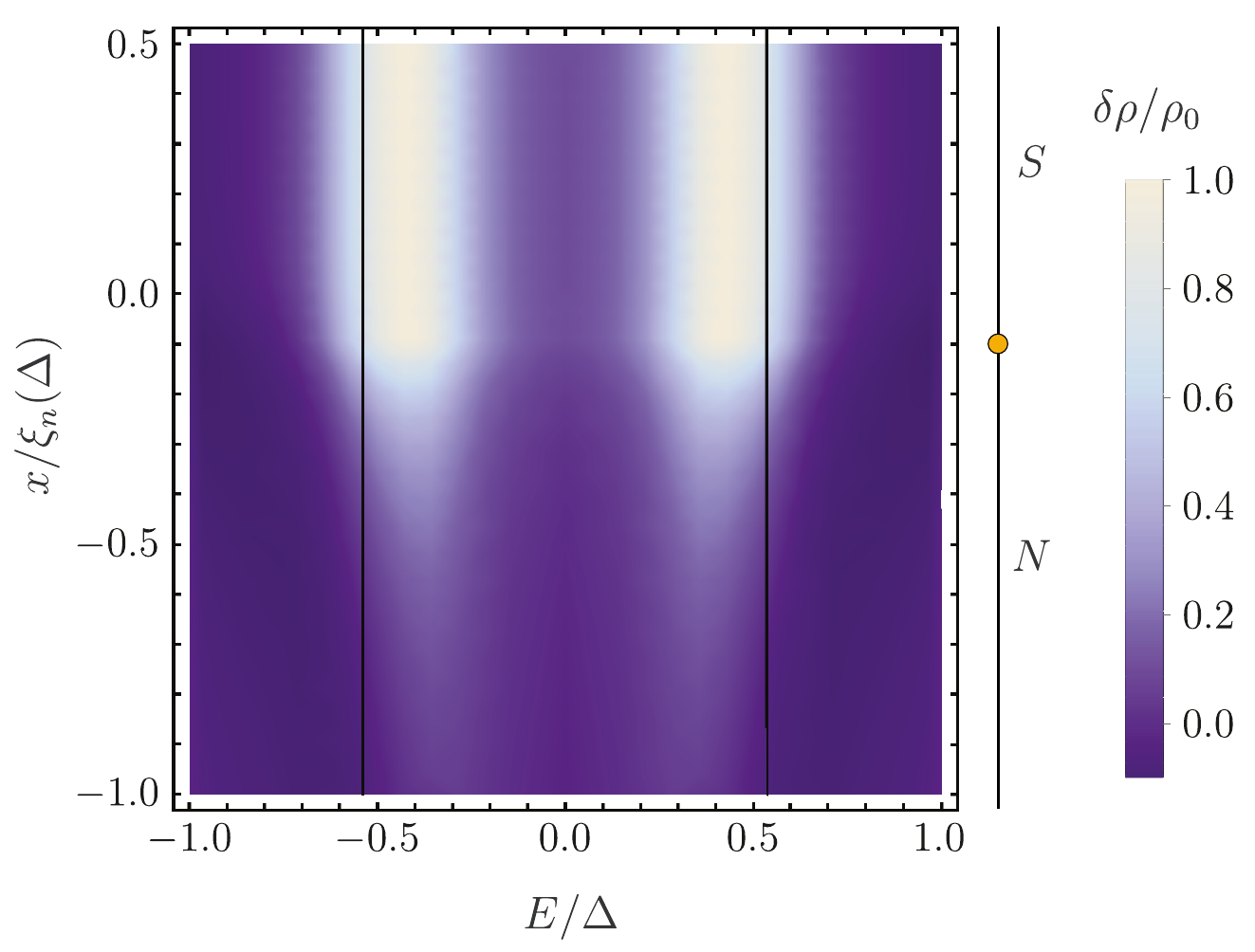}}
\caption{Dependence of the relative change in the LDoS $\delta\rho{/}\rho_0$ on energy and $x$-coordinate as a result of the magnetic impurities located at $x{=}{-}b{=}{-}0.1\,\xi_{\rm n}(\Delta)$. Black vertical lines denote the position of the YSR energy ($\alpha{=}0.3$). We use $n_s\xi(0)/g=0.002$, 
$\sqrt{D_{\rm n}/D}{=}20$ and $g{=}g_{\rm n}$.}
\label{Fig:figure8}
\end{figure} 

Fig. \ref{Fig:figure7} shows the energy dependence of the LDoS for different values of $b$. Again, the peak is displaced from the YSR energy (see Eq. \eqref{III:eq:AlgebraicSolve:N2}). As can be seen, when the impurity is removed farther away from the SN boundary, the peak is blurred, becoming lower and wider. This behavior is in agreement with the inequality \eqref{III:eq:Condition:3}.

Fig. \ref{Fig:figure8} displays the energy and coordinate dependence of the relative size of the shifted peak with distance from the impurity. As in Fig. \ref{Fig:figure5}, in the superconductor region, the ratio $\delta\rho/\rho_0$ does not depend on the coordinate.

As one can see from Figs. \ref{Fig:figure7} and \ref{Fig:figure8}, the LDoS for magnetic impurities situated in the normal metal within the distance $b{\ll}\xi_{\rm n}$ is qualitatively the same as the one in the case $b{=}0$. 

 For completeness, in Appendix \ref{App:Impurity} we present the density of states in the case of SN junction with a chain of magnetic impurities situated in the superconducting region.
\color{black} 

\section{Discussions and conclusions\label{Sec:3}}

In our paper, we do not take into account the spin-independent part of the magnetic impurity  strength, $\alpha_0$. However, its effect can be incorporated into redefinition of the parameter $\beta {\to} (1{-}\alpha{+}\alpha_0)^2{/}(4\alpha)$. Therefore, all our results can be easily applied to that more general case.
 
The results of Sec. \ref{Sec:1:Stand} for a solitary magnetic impurity 
 are \color{black} related with the case of finite impurity concentration $n_s^{(2)}$. We remind that a single magnetic impurity produces a perturbation of the LDoS with spatial extent of the order of the superconducting coherence length at the YSR energy, $\xi_\beta$. Therefore, we can expect our results to be applicable for a finite impurity concentration $n_s^{(2)}{\sim}1/\xi_\beta^2$. In this case, using Eq. (96) of Ref. \cite{FS2016}, we find the width of the impurity band to be of the order of $\Delta/[(1+\beta)\sqrt{g}]$. The latter estimate coincides with $\Gamma$ up to a logarithm in the definition of spreading resistance $t$, cf. Eq. \eqref{eq:def:spreading resistance}.  We emphasize that two seemingly different problems -- spatially inhomogeneous one for a solitary magnetic impurity and homogeneous one for a finite concentration of magnetic impurities -- occur to be related.   
\color{black}

As mentioned above, the broadening of the LDoS near the YSR energy is caused by the  fluctuations  of the dimensionless effective strength $\alpha$ of a magnetic impurity. Therefore, it would be tempting to say that the distribution of the YSR energy (defined as the energy at which the peak in the LDoS has the maximum) can be directly read from the log-normal distribution \eqref{eq:P:dist} and Eq. 
\eqref{eq:YSR:energy}. However, Eq. \eqref{eq:EYSR:mod:UE} demonstrates clearly that this is not the case. In fact, the problem of computation of the YSR energy distribution in a dirty superconducting film is more complicated and goes far beyond the present work.

We emphasize that in the case of a single magnetic impurity randomness of the YSR state is introduced due to different realizations of potential disorder. It should be contrasted with the case of rare magnetic impurities considered in Ref. \cite{BurmistrovSkvortsov2018} where fluctuations of YSR states at different magnetic impurities were related to the point-to-point fluctuations of the local density of states due to potential disorder. Surprisingly, on the level of the Usadel equation both effects can be described by the very same log-normal distribution of the dimensionless effective strength $\alpha$, cf. Eq. \eqref{eq:P:dist}.     
\color{black}

In our work, in order to find the LDoS, we solve the Usadel equation for a spatially dependent spectral angle. We remind that the Usadel equation corresponds to the saddle-point treatment of the NLSM (see Refs. \cite{MaS,FS2016,BurmistrovSkvortsov2018} for details). Renormalization of the NLSM action between the length scales $\ell$ and $\xi$ should be taken into account. It leads to the renormalized Usadel equation. However, one can treat the renormalized NLSM beyond the saddle-point approximation. This results in additional fluctuation corrections to the LDoS. For a finite impurity concentration in a dirty  superconducting film, one can estimate the relative fluctuation correction to the LDoS to be of the order of ${\sim}(n_s^{(2)}\xi^2)/g^2$ \cite{Pashinsky2021}. 
Applying this estimate with $n_s^{(2)}{\sim}1/\xi^2$ for a solitary magnetic impurity, we find that the fluctuation corrections to the LDoS are negligible in comparison with the results derived from the Usadel equation. We expect a similar conclusion in the case of a magnetic impurity chain near the SN boundary.

In our paper, we treat the magnetic impurity spin fully classically. Such approximation is formally justified by the limit $S{\gg}1$. Since in reality the magnetic impurity spin is not that large, $S{\lesssim}5/2$, it would be interesting to investigate the effect of potential disorder on the YSR state treating the spin quantum mechanically (for a clean case see Ref. \cite{Oppen2021} and references therein).
	
For a chain of magnetic impurities, the quantum dynamics of their spins leads to an intriguing competition between the
Kondo effect and the indirect exchange interaction that can be probed by STM measurements \cite{Franke2021o,Franke2021e}. Taking into account potential disorder is likely to be important for interpretation of the STM data. 

For a chain of rare magnetic impurities near the SN boundary,  we limit our consideration by simplest geometry when the chain is parallel to the SN interface. Recently, YSR-type features in the LDoS at grain boundaries in graphene with Pb islands have been measured \cite{Rio2021}. In view of these experimental findings, it would be worthwhile to study more complicated geometries of magnetic impurity chains near SN interfaces.

To summarize, we reported the results of detailed studies of the effects of potential disorder on YSR states in superconducting films. We focus on two setups: (i) a solitary magnetic impurity in a dirty superconducting film and (ii) a chain  of magnetic impurities situated in a normal region of an SN junction. Solving the Usadel equation for a spatially dependent spectral angle, we found that potential disorder broadens the YSR state. This manifests as the peak in LDoS at energies near $E_{\rm YSR}$. 

The    broadening of the peak is proportional to the square root of resistance per square of the film. Thus, it is larger than one could naively expect. The physical mechanism for appearance of broadening is fluctuations of the LDoS in the normal state. The latter results in 
fluctuations of dimensionless impurity strength $\alpha$ and, consequently, to fluctuations of an energy of the YSR state. In the case of a single magnetic impurity in a dirty superconducting film, we demonstrate that 

	modification of multiple scattering on the magnetic impurity due to intermediate scattering on surrounding potential disorder \color{black}
 is of crucial importance for correct description of the LDoS profile near the YSR energy. In particular, the account of 
  this modification \color{black} allowed us to remove unphysical abrupt vanishing of the LDoS obtained within standard Usadel equation. We are not aware of any systematic experimental studies of the dependence of the YSR peak width in the LDoS on the sheet resistance of a film. 

We demonstrated that existence of a normal metal makes the YSR state to be the quasibound state rather than the bound one. For a solitary magnetic impurity in a dirty superconducting film, such an effect is caused by the normal-metal tip used for STM measurements. In the case of a magnetic impurity chain, the normal region of the SN heterostructure provides a channel for decay of the YSR state. 

Finally, we mention that it would be interesting to extend our study to superconducting systems with spin-orbit coupling in which a magnetic impurity chain can host Majorana bound states together with YSR states.

\begin{acknowledgements}
		
The authors are grateful to Ya. Fominov, A. Melnikov, M. Skvortsov for very useful discussions. We are especially grateful to I. Tamir for providing us the experimental data on the LDoS. The research was partially supported by the Russian Ministry of Science and Higher Education, the Russian Foundation for Basic Research (grant No. 20-52-12013) - Deutsche Forschungsgemeinschaft (grant No. EV 30/14-1) cooperation, and by the Basic Research Program of HSE. A. Lyublinskaya is also grateful to JetBrains Co. Ltd. for a personal scholarship through the program to support women and girls in STEM. 
	\end{acknowledgements}
	
\appendix

\section{Derivation of Eq. \eqref{eq:eq:tilde:psi} from the standard Usadel equation \eqref{eq:Usadel:1} \label{App:Self-consistent}}

\noindent
In this Appendix we present brief derivation of Eq. \eqref{eq:eq:tilde:psi} from the standard Usadel equation \eqref{eq:Usadel:1}. It is expressed as follows 
\begin{align}
 & \xi^{2}\nabla^{2}\delta\psi_{\sigma}-\sinh\delta\psi_{\sigma}=\frac{\left[\xi^{2}/(\pi\nu D)\right]\cosh\psi_{\sigma}}{\sigma\sqrt{\beta}+\sinh\psi_{\sigma}}\delta(\boldsymbol{r}) .
\end{align}
\noindent
Taking into account the smallness of deviation from the homogeneous solution,  $\left|\delta\psi_{\sigma}\right|\ll1$, 
we can treat the linearized equation,
\begin{align}
 & \xi^{2}\nabla^{2}\delta\psi_{\sigma}-\delta\psi_{\sigma}=\frac{\left[\xi^{2}/(\pi\nu D)\right]\cosh\psi_{\sigma}}{\sigma\sqrt{\beta}+\sinh\psi_{\sigma}}\delta(\boldsymbol{r}) .
 \label{eq:1:App:A}
\end{align}
This equation for $\delta\psi_{\sigma}$ is similar to the 2D Schr\"odinger equation with $\delta(\bm{r})$ potential. For $r{>}l$ the solution of Eq. \eqref{eq:1:App:A} can be written as
\begin{equation}
 \delta\psi_{\sigma}\left(r\right)=\left(\tilde{\psi}_{\sigma}-\psi_{\infty}\right)\cdot\frac{K_{0}\left(r/\xi\right)}{\ln\xi/l} .
\end{equation}
\noindent
Here we have  introduced the notation $\tilde{\psi}_{\sigma}=\delta\psi_{\sigma}\left(l\right)+\psi_{\infty}$
and have taken into account the small parameter  $l/\xi\ll1$.

In order to treat the delta-functional potential we apply the Fourier transformation
to the equation \eqref{eq:1:App:A},
\begin{equation}
 -\left(q^{2}\xi^{2}+1\right)\delta\psi_{\sigma}\left(\bm{q}\right)=\frac{\left[\xi^{2}/(\pi\nu D)\right]\cosh\tilde{\psi}_{\sigma}}{\sigma\sqrt{\beta}+\sinh\tilde{\psi}_{\sigma}} .
\end{equation}
Here $\delta\psi_{\sigma}\left(\bm{q}\right)$ is the Fourier transform of $\delta\psi_{\sigma}\left(r\right)$. Thus, we derive the self-consistent equation \eqref{eq:eq:tilde:psi} for $\tilde{\psi}_{\sigma}$.

\color{black}


\section{Derivation of the Usadel equation for a solitary magnetic impurity\label{App:Usadel}} 	

The NLSM action for a dirty superconducting film with a solitary magnetic impurity can be written as (see Ref. \cite{Burmistrov2021} for details):
\begin{gather}
S=S_\sigma + S_{\Delta}+S_{\rm mag} .
\label{app:eq:NLSM}
\end{gather}
Here the first term in the right-hand side of Eq. \eqref{app:eq:NLSM} is given by
\begin{equation}
S_\sigma  = \frac{g}{32} \int d\bm{r} \Tr (\nabla Q)^2 - 2 Z_\omega \int d\bm{r} \Tr [\hat \varepsilon +\hat \Delta] Q  . \label{app:Ss}
\end{equation}
The field $Q(\bm{r})$ is a matrix in the replica, spin, Matsubara, and particle-hole spaces. The trace $\Tr$ acts in the same spaces. The matrix field $Q$ obeys the nonlinear constraint and charge-conjugation symmetry relation,
\begin{gather}
Q^2(\bm{r})=1, \quad \Tr Q = 0, \quad  Q=Q^\dag = -C Q^T C,
\label{app:eq:constraints}
\end{gather}
where $C=i t_{12}$. The action \eqref{app:Ss} involves two constant matrices:
\begin{equation}
\hat \varepsilon_{nm}^{\alpha\beta} =\varepsilon_n \, \delta_{\varepsilon_n,\varepsilon_m}\delta^{\alpha\beta} t_{00}, \quad 
\hat\Delta_{nm}^{\alpha\beta}  =  \Delta \delta_{\varepsilon_n,-\varepsilon_m} \delta^{\alpha\beta}t_{10} .
\end{equation}
Here $\alpha,\beta {=} 1,\dots, N_r$ stand for replica indices, while integers $n,m$ correspond to the Matsubara fermionic frequencies $\varepsilon_n {=} \pi T (2n{+}1)$. The superconducting order parameter $\Delta$ is assumed to be a real scalar. The sixteen matrices,
\begin{equation}
\label{trj}
t_{rj} = \tau_r\otimes s_j, \qquad r,j = 0,1,2,3  ,
\end{equation}
operate in the spin (subscript $j$) and particle-hole (subscript $r$) spaces. The matrices 
$\tau_r$ and $s_r$ are the standard Pauli matrices. 
We note that the parameter $Z_\omega$ describes the frequency renormalization upon the renormalization group flow (see Ref. \cite{Fin} for details). The bare value of $Z_\omega$ is equal to $\pi \nu/4$.
The second term of the action \eqref{app:eq:NLSM} reads
\begin{equation}
S_{\Delta} = - \frac{4 Z_\omega N_r}{\pi T \gamma_{c0}} \int d\bm{r} \Delta^2 .
\label{app:Sd}
\end{equation}
The last term of $S$ describes the action of the solitary magnetic impurity,
\begin{equation}
S_{\rm mag} = - \frac{1}{2}\Tr \ln \bigl (1+i\sqrt{\alpha_0} Q(0) + i \sqrt{\alpha} Q(0) t_{33} \bigr ) .  
\end{equation}

We choose the following form of the saddle-point $Q$-matrix,
\begin{align}
\underline{Q}_{nm}^{\alpha\beta} = &  \frac{1}{2}\sum_{\sigma=\pm}\Bigl (\cos\theta_{\sigma}
(t_{00} \sgn \varepsilon_n +\sigma t_{33}) \delta_{\varepsilon_n,\varepsilon_m}
\notag \\
 & + \sin \theta_{\sigma}
 ( t_{10}-i \sigma t_{23} \sgn \varepsilon_n )\delta_{\varepsilon_n,-\varepsilon_m} \Bigr )\  \delta^{\alpha\beta} .
 \label{app:eq:sol:sp}
\end{align}
Here we assume that the spectral angle $\theta_{\sigma}{\equiv}\theta_{\sigma}(\varepsilon_n)$ is an even function of $\varepsilon_n$. Then variation of the saddle-point action $S[\underline{Q}]$ with respect to the spectral angle $\theta_{\sigma}(\varepsilon_n)$ 
results in the standard Usadel equation \eqref{eq:Usadel:1}. Varying $S[\underline{Q}]$ over $\Delta$ yields a self-consistent equation for the superconducting order parameter, Eq. \eqref{eq:self-consist:1}. 

In order to derive the renormalized Usadel equation, we need to consider the renormalization of the NLSM action. Let us split the matrix field $Q$ into the fast $q$ and slow $Q_0{=}T^{-1}\Lambda T$ components. Here we introduce the matrix 
\begin{equation}
\Lambda_{nm}^{\alpha\beta} =\sgn \varepsilon_n \, \delta_{\varepsilon_n,\varepsilon_m}\delta^{\alpha\beta} t_{00} .
\end{equation}
The renormalized action for a magnetic impurity is determined as follows,
\begin{equation}
S_{\rm mag}^{\rm (ren)}[Q_0]=
- \ln \left \langle e^{-S_{\rm mag}[T^{-1} q T]}\right \rangle_{q} .
\end{equation}
Here the averaging $\langle\dots\rangle_{q}$ is with respect to the NLSM action $S_{\sigma}$ for the fast modes $q$.    
As was derived in Ref. \cite{BurmistrovSkvortsov2018}, 
the term $\exp(-S_{\rm mag}[T^{-1} q T])$ transforms upon renormalization as follows,
\begin{gather}
\left \langle e^{\frac{1}{2} \Tr\ln (1+i\sqrt{\alpha} T^{-1} q(0) T t_{33})}\right \rangle_q {\to} \left \langle e^{\frac{1}{2} \Tr\ln (1+i\sqrt{\textrm{a}} Q_0(0) t_{33})}\right \rangle_{\textrm{a}} .
\end{gather}
Here we set $\alpha_0{=}0$ for a sake of simplicity. The average $\langle \dots\rangle_{\textrm{a}}$ is defined with respect to the distribution function \eqref{eq:P:dist}. 
For the derivation of the Usadel equation we need to know the saddle-point action in the replica limit, $N_r{\to}0$, alone. Therefore, we find
 \begin{equation}
 S_{\rm mag}^{\rm (ren)}[\underline{Q}] \simeq - \frac{1}{2} \Bigl \langle \tr\ln \bigl (1+i\sqrt{\textrm{a}} \underline{Q}(0) t_{33}\bigr )\Bigr \rangle_{\textrm{a}} .
 \end{equation}
Varying $S_{\sigma}[\underline{Q}]{+}S_{\rm mag}^{\rm (ren)}[\underline{Q}]$ over the spectral angle $\theta_{\sigma}(\varepsilon_n)$ yields the renormalized Usadel equation \eqref{eq:Usadel:1:mod}. 

We note that for a nonzero $\alpha_0$, we would obtain the renormalized Usadel equation with the distribution functions \eqref{eq:P:dist} for quantities corresponding to both  $(\sqrt{\alpha}{+} \sqrt{\alpha_0})^2$ and $(\sqrt{\alpha}{-} \sqrt{\alpha_0})^2$.



\section{Condition for  rareness of magnetic impurities in the case of SN junction\label{App:Condition}}	
	
The MLSM approach allows us to establish the condition of rareness of magnetic impurities. In the case of a superconducting film, the corresponding  condition can be formulated as  \cite{BurmistrovSkvortsov2018}
\begin{equation}
\frac{n_s}{\nu_{\rm n}} |\mathcal{D}(\bm{r},\bm{r})| \gg \frac{n_s^2}{\nu_{\rm n}^2} \int d^2\bm{r^\prime}
|\mathcal{D}(\bm{r},\bm{r}^\prime) \mathcal{D}(\bm{r}^\prime,\bm{r})| ,
\label{app:cond:DD}
\end{equation}
where $\mathcal{D}(\bm{r},\bm{r}^\prime)$ stands for the diffusion propagator. In the case of a homogeneous 2D superconductor, the diffusion propagator can be written as
\begin{equation}
\mathcal{D}(\bm{r},\bm{r}^\prime) = \int \frac{d^2\bm{q}}{(2\pi)^2} \frac{e^{i\bm{q}(\bm{r}-\bm{r^\prime})}}{D(q^2+\xi^{-2})} .
\end{equation} 
Hence we find the inequality \eqref{app:cond:DD} reduces to the condition, $n_s\xi^2/g {\ll} 1$. We note that we neglect a logarithmic factor.  

In the case of a chain of magnetic impurities parallel to the SN boundary, one needs to find the diffusive propagator. It satisfies the following equations:
\begin{gather}
D_{\rm n} [-\partial_x^2 +q_y^2-i \xi_{\rm n}^{-2}] \mathcal{D}(q_y;x,x^\prime)=\delta(x-x^\prime), \quad x<0 \notag  \\
D [-\partial_x^2 +q_y^2+\xi^{-2}] \mathcal{D}(q_y;x,x^\prime)=\delta(x-x^\prime), \quad x>0 .
\label{app:eq:DD}
\end{gather}	
The boundary condition at $x{=}0$ reads
\begin{gather}
\mathcal{D}(q_y;-0^+,x^\prime)=\mathcal{D}(q_y;0^+,x^\prime), \notag \\
g_{\rm n} \partial_x \mathcal{D}(q_y;x,x^\prime)|_{x=-0^+}=
g \partial_x \mathcal{D}(q_y;x,x^\prime)|_{x=0^+} . 
\end{gather}
Here we perform the Fourier transform with respect to $y$ coordinate (which is parallel to the SN boundary). Hence, in the case of a magnetic impurities chain, the condition \eqref{app:cond:DD} becomes
 \begin{gather}
\int \frac{dq_y}{2\pi}|\mathcal{D}(q_y;-b,-b)| \gg \frac{n_s}{\nu_{\rm n}} \int \frac{d q_y}{2\pi}
|\mathcal{D}(q_y;-b,-b)|^2 ,
\label{app:cond:DD:1}
\end{gather}

\begin{figure*}[t]
\includegraphics[width=0.47\textwidth]{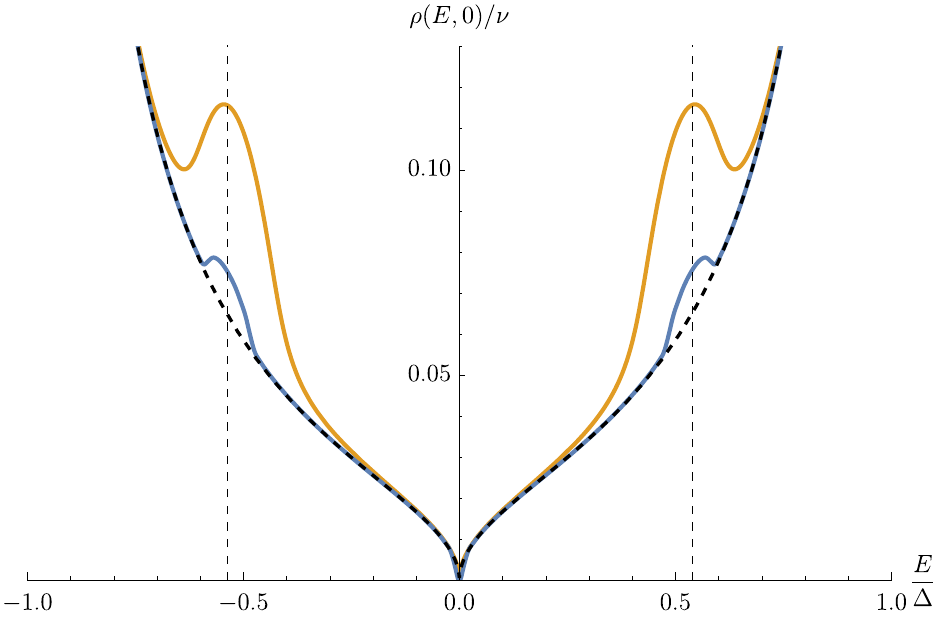}\qquad
\includegraphics[width=0.47\textwidth]{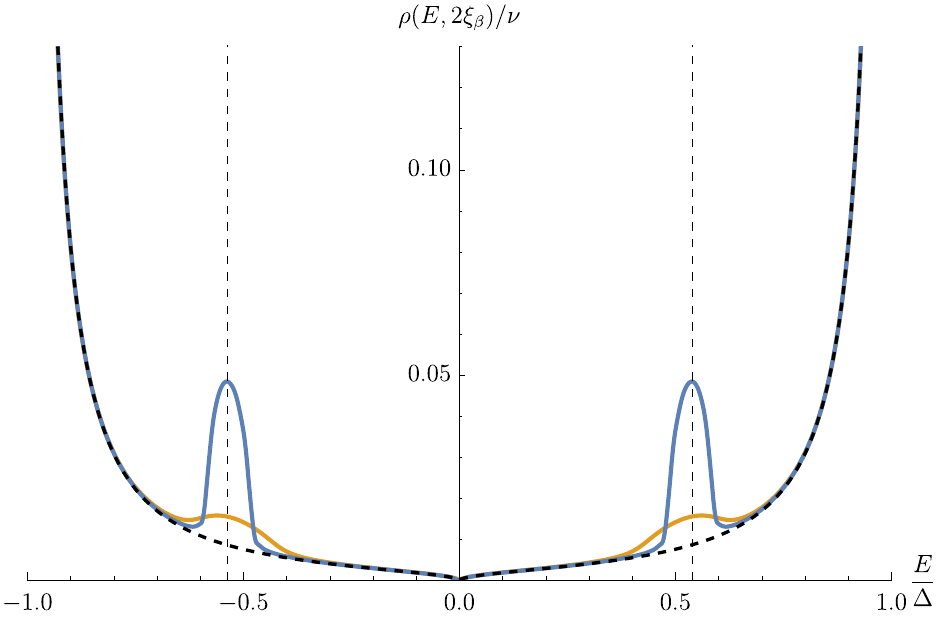}
\caption{Dependence of the LDoS on the energy at the SN boundary $x{=}0$ (left), and at the point $x{=}2\xi_{\beta}$ (right). Vertical dashed lines denote the position of the YSR energy at a given value of $\alpha{=}0.3$. Orange and blue curves show the LDoS for the impurities situated at $b{=}0$ and at $b{=}2\xi_{\beta}$, respectively. The dashed curves denote, for comparison, the LDoS without magnetic impurities, i.e., for $n_s{=}0$. We use $n_s\xi(0)/g=0.002$, 
$\sqrt{D_{\rm n}/D}{=}20$ and $g{=}g_{\rm n}$.}
\label{Fig:figure9}
\end{figure*} 

Solving Eqs. \eqref{app:eq:DD}, we find the following expression for the diffusive propagator,
\begin{gather}
\mathcal{D}(q_y;-b,-b) = \frac{1}{2D_{\rm n}\sqrt{q_y^2-i\xi_{\rm n}^{-2}}}
\Biggl [ 1 - \frac{\gamma \frac{\sqrt{1+q_y^2\xi^2}}{\sqrt{1+i q_y^2\xi_{\rm n}^2}} -e^{-\frac{i\pi}{4}} }{\gamma \frac{\sqrt{1+q_y^2\xi^2}}{\sqrt{1+i q_y^2\xi_{\rm n}^2}} +e^{-\frac{i\pi}{4}}} \notag \\
\times e^{-2 b\sqrt{q_y^2-i\xi_{\rm n}^{-2}}}
\Biggr ] .
\end{gather}
In the case $b{\gg}\xi_{\rm n}$, the inequality \eqref{app:cond:DD:1} reduces to the following condition, 
\begin{equation}
n_s \xi_{\rm n}/g_{\rm n} \ll 1 .
\label{app:ineq:1}
\end{equation}
In the opposite case, $b{\ll}\min\{\xi_{\rm n},\xi\}$, we find from Eq. \eqref{app:cond:DD:1}, the following inequalities
\begin{equation}
\begin{split}
{n_s \xi}/{g} \ll 1 , \qquad \gamma\gg 1 ,  \\
{n_s \xi_{\rm n}}/{\max\{g,g_{\rm n}\}} \ll 1, \qquad \gamma\ll 1 .
\end{split}
\label{app:ineq:2}
\end{equation}
As one can see, Eqs. \eqref{app:ineq:1} and  \eqref{app:ineq:2} are equivalent to Eq.~\eqref{eq:ns:cond}.

\section{YSR resonance in a dirty SN
junction with magnetic impurities situated inside the superconductor. \label{App:Impurity}}

In this Appendix we consider how a chain of magnetic impurities situated inside the superconducting region in SN junction affects the density of states. We shall perform  calculations in a way similar to the one described in Sec. \ref{III:standard}. We assume that the chain is parallel to the SN interface and is situated at the point $x{=}b$. By analogy with Eqs. \eqref{III:eq:UsadelSolve:1a}-\eqref{III:eq:UsadelSolve:1}, we write out the spectral angle in the region $x{<}0$,
\begin{equation}
\theta_{\rm n}(E,x)=4\arctan\left [ \exp \left (\frac{x}{\xi_{\rm n}}e^{-i\pi \sgn E/4}\right ) \tan\frac{\theta_{\rm 0}}{4}\right],
\label{App:eq:UsadelSolve:1}
\end{equation}
and the region $x{>}b$, 
\begin{equation}
\theta_{\rm s}(E,x)=\frac{\pi}{2}+i\psi_{\infty}+4i\arctanh\left(e^{-(x-b)/\xi}\tanh\frac{\psi_{\rm b}}{4}\right).
\label{App:eq:UsadelSolve:2}
\end{equation}

Next, to find a solution on the interval $0{<}x{<}b$, we use the first integral of the Usadel equation:
\begin{equation}
\frac{\xi^2}{2}\left(\partial_x\theta_{\rm s}\right)^2 + \sin\left(\theta_{\rm s}-i\psi_{\infty}\right) = C,
\end{equation}
where $C$ is the constant. This allows us to reduce the solution to the inversion of the incomplete elliptic integral,
\begin{equation}
\frac{b-x}{\xi}=\int\limits_{\frac{\pi}{2}+i\psi_{\infty}+i\psi_{\rm b}}^{\theta_{\rm s}(E,x)}\frac{d\theta}{\sqrt{2C-2\sin\left(\theta-i\psi_{\infty}\right)}}.
\label{App:eq:UsadelInt:2}
\end{equation}

To fully determine the spectral angle, we need to find constants $\theta_0$, $\psi_{\rm b}$, and $C$. Boundary conditions \eqref{III:eq:stitching:1} at the point $x{=}0$ yield
\begin{equation}
C=\frac{2g_{\rm n}^2\xi^2}{g^2\xi_{\rm n}^2}e^{-i\pi\sgn E/2}\sin^2\frac{\theta_{\rm0}}{2}+\sin\left(\theta_{\rm0}-i\psi_{\infty}\right)
\label{App:eq:algebraic:1}
\end{equation}
and at the point $x{=}b$ lead to 
\begin{gather}
C=-2\left(\sinh\frac{\psi_{\rm b}}{2}-\frac{(4 n_s\xi\alpha/g) \sinh(2\psi_{\infty}+2\psi_{\rm b})}{1+\alpha^2 -2 \alpha \cosh(2\psi_{\infty}+2\psi_{\rm b})}\right)^2\notag \\
+ \cosh\psi_{\rm b}.
\label{App:eq:algebraic:2}
\end{gather}

The last equation is obtained from the Eq. \eqref{App:eq:UsadelInt:2} by substituting $x{=}0$:
\begin{equation}
\frac{b}{\xi}=\int\limits_{\frac{\pi}{2}+i\psi_{\infty}+i\psi_{\rm b}}^{\theta_{\rm 0}}\frac{d\theta}{\sqrt{2C-2\sin\left(\theta-i\psi_{\infty}\right)}}.
\label{App:eq:algebraic:3}
\end{equation}

Thus, by substituting the constants $\theta_0$, $\psi_{\rm b}$, and $C$ obtained from the solution of the algebraic system \eqref{App:eq:algebraic:1}-\eqref{App:eq:algebraic:3} into the equations \eqref{App:eq:UsadelSolve:1}, \eqref{App:eq:UsadelSolve:2}, \eqref{App:eq:UsadelInt:2}, we completely determine the spectral angle.

Using the obtained expressions, one can find the dependence of the local density of states on the energy numerically. On the Fig. \ref{Fig:figure9}, we show dependence of the density of states on energy at $x{=}2\xi_{\beta}$ for two positions of the impurity chain: at $b{=}0$ and $b{=}2\xi_{\beta}$.

\color{black}

\bibliography{literature_MagImp}	
	
\end{document}